\documentclass[12pt]{article}
\usepackage{cite}
\usepackage{pifont}
\usepackage{color}
\usepackage{amsmath}
\usepackage{hyperref}
\usepackage{graphicx}
\graphicspath{{./}{./figures/}}

\newcommand{\vep}{\varepsilon}
\newcommand{\ep}{\epsilon}
\newcommand{\MSbar}{\overline{\mbox{MS}}}
\newcommand{\msbar}{\overline{\mbox{\scriptsize MS}}}
\newcommand{\ARI}{\mbox{RI/MOM}}
\newcommand{\ari}{\mbox{\scriptsize RI/MOM}}
\newcommand{\RIp}{\mbox{RI}^\prime\mbox{/MOM}}
\newcommand{\rip}{\mbox{\scriptsize RI}^\prime\mbox{\scriptsize/MOM}}
\newcommand{\RISMOM}{\mbox{RI/SMOM}}
\newcommand{\RISMOMB}{\mbox{RI/SMOM}_{\gamma_{\mu}}}
\newcommand{\rismomb}{{\scriptsize{\mbox{RI/SMOM}_{\gamma_{\mu}}}}}
\newcommand{\rismom}{\mbox{\scriptsize RI/SMOM}}

\newcommand{\Tr}{\mbox{Tr}}
\newcommand{\Fslash}[1]{\!\not{\hbox{\kern-2pt ${#1}$}}}
\newcommand{\unitop}{1{\hbox{\kern-7pt $1$}}}
\newcommand{\CF}{C_F}
\newcommand{\CA}{C_A}
\newcommand{\TF}{T_F}
\newcommand{\nf}{n_f}
\newcommand{\M}[3]{{\mathcal{M}}^{(#1)}_{#2#3}}
\newcommand{\Order}[1]{{\mathcal{O}}{(#1)}}
\newcommand{\nlog}[1]{\log\left(#1\right)}
\newcommand{\Log}[2]{\log^#2\!\left(#1\right)}
\newcommand{\z}[1]{\zeta_{#1}}
\newcommand{\s}[2]{s_#1\!\left(\tfrac{\pi}{#2}\right)}

\newcommand{\p}[1]{\Psi'\!\left(\tfrac{1}{#1}\right)}
\newcommand{\ppp}[1]{\Psi'''\!\left(\tfrac{1}{#1}\right)}
\allowdisplaybreaks

\begin{document}    
\begin{titlepage}
\noindent
\mbox{}\hfill \\
\mbox{}\hfill{\bf{YITP-SB-1015}}
%\mbox{}\hfill{\bf{ Date: \today}}
%

\vspace{0.5cm}
\begin{center}
  \begin{Large}
    \begin{bf}
      \begin{center}
%      Title
%%%%%%%%%
      Two-loop matching factors for light quark masses and 
      three-loop mass anomalous dimensions in the $\RISMOM$ schemes  
%%%%%%%%%
      \end{center}
    \end{bf}
  \end{Large}
  \vspace{0.8cm}

  \begin{large}
  \end{large}
  \vskip .7cm
	   Leandro G. Almeida $\rm ^{a,b}$ and
	   Christian Sturm $\rm ^{b}$ \\
	   \vspace{1cm}
    {\small {\em $\rm ^a$  
                  C.N. Yang Institute for Theoretical Physics, 
                  Stony Brook University,\\
                  Stony Brook, New York 11794, USA}}\\
  {\small {\em $\rm ^b$
	     Physics Department, 
             Brookhaven National Laboratory, \\
             Upton,
	     New York 11973, USA}}\\
	\vspace{2cm}
{\bf Abstract}
\end{center}
\begin{quotation}
  \noindent
Light quark masses can be determined through lattice simulations in
regularization invariant momentum-subtraction($\ARI$)
schemes. Subsequently, matching factors, computed in continuum
perturbation theory, are used in order to convert these quark masses
from a $\ARI$ scheme to the $\MSbar$ scheme. We calculate the two-loop
corrections in quantum chromodynamics(QCD) to these matching factors as
well as the three-loop mass anomalous dimensions for the $\RISMOM$ and
$\RISMOMB$ schemes. These two schemes are characterized by a symmetric
subtraction point. Providing the conversion factors in the two different
schemes allows for a better understanding of the systematic
uncertainties. The two-loop expansion coefficients of the matching
factors for both schemes turn out to be small compared to the
traditional $\ARI$ schemes. For $\nf=3$ quark flavors they are about
0.6\%-0.7\% and 2\%, respectively, of the leading order result at scales of
about 2~GeV. Therefore, they will allow for a significant reduction of
the systematic uncertainty of light quark mass determinations obtained
through this approach. The determination of these matching factors
requires the computation of amputated Green's functions with the
insertions of quark bilinear operators. As a by-product of our
calculation we also provide the corresponding results for the tensor
operator.
\end{quotation}
\end{titlepage}
\section{Introduction}
\label{sec:Introduction}
Light quark masses, like the up-, down-, and strange-quark masses, are
fundamental parameters of the Standard Model, and their precise
determination is thus an important task. They can be determined, for
example, with the help of lattice simulations in combination with
nonperturbative renormalization(NPR). In this context regularization
invariant momentum-subtraction ($\ARI$) schemes\cite{Martinelli:1994ty}
play a crucial role; for a recent overview see e.g. Ref.~\cite{Aoki:2010yq}.

In Ref.~\cite{Allton:2008pn} these light quark masses were determined in
the $\ARI$ scheme and subsequently converted to the $\MSbar$
scheme~\cite{tHooft:1973mm,Bardeen:1978yd}. This conversion requires the
computation of a matching factor $C^x_m$, which performs this
transformation of the quark mass from the scheme $x$ into the $\MSbar$
mass. Since the $\ARI$ schemes do not depend on the particular regulator
which has been used to regularize the ultraviolet divergences, this
matching factor can be calculated in continuum perturbation theory using
dimensional regularization\cite{tHooft:1972fi}. In the $\ARI$ scheme
this matching factor is known up to three-loop order in perturbative
QCD~\cite{Martinelli:1994ty,Franco:1998bm,Chetyrkin:1999pq}; the same
holds for the $\RIp$~\cite{Chetyrkin:1999pq,Gracey:2003yr} scheme.

Both the $\ARI$ and $\RIp$ schemes employ a renormalization procedure
which uses an exceptional subtraction point. Here the subtraction point
defines the configuration of the external momenta of the considered
amplitude, where the ultraviolet divergences are subtracted.  In
continuum perturbation theory, mass and fermion field renormalization
constants are typically computed by considering higher order corrections
to fermion self-energy diagrams. The fermion self-energies are related
through Ward-Takahashi identities to amputated Green's functions of
quark bilinear operators, {\it i.e.} the vector, axial-vector, scalar and
pseudoscalar operators. These relations allow an extraction of the 
renormalization constants from these amputated Green's functions with
operator insertions, rather than from self-energies. In the case of an
exceptional momentum configuration, no momentum transfer leaves the
operator.

However, at low scales ($\sim$ 2 GeV) the perturbative expansion of the
matching factors for the $\ARI$ and $\RIp$ schemes exhibit a poor
convergence behavior and, as a result, introduce large systematic
uncertainties in the determination of the $\MSbar$ masses of the light
quark sector. This amounts to approximately 60\% of the total error.
Apart from that, the lattice simulations are, for an exceptional
subtraction point, more prone to unwanted infrared effects. For this
reason, the use of a symmetric subtraction point was proposed in
Ref.~\cite{Aoki:2007xm}, and the concepts and framework of these new
$\RISMOM$ schemes have been worked out in Ref.~\cite{Sturm:2009kb}. A
nonexceptional or symmetric subtraction point is characterized by the
fact that a momentum leaves the operator of the amputated Green's
function, and in the case of a symmetric subtraction point, the squares of
all momenta leaving the amplitude are equal. A nonperturbative test of
such a $\RISMOM$ scheme can be found in Ref.~\cite{Aoki:lat2008}.  The
one-loop QCD corrections of these matching factors in the $\RISMOM$
schemes are known~\cite{Sturm:2009kb}, and are shown to have a better
convergence behavior than the traditional $\ARI$ and $\RIp$
schemes. However, the question of whether this behavior persists at
higher orders in perturbation theory still remained unanswered. If it
were confirmed, this would lead to a significant reduction of the
systematic uncertainties associated with the determination of the light
quark masses.

The goal of this paper is to extend the work of
Ref.~\cite{Sturm:2009kb} and to provide the two-loop QCD corrections to
the matching factors of the so-called $\RISMOM$ and $\RISMOMB$ schemes
as well as the three-loop mass anomalous dimensions.

This paper is structured as follows: In Section~\ref{sec:Generalities}
we introduce the notations and conventions that are used. In
Section~\ref{sec:Calculation} we give an outline of the perturbative
calculation, and in Section~\ref{sec:Results} we present the results for
the matching factors and the mass anomalous dimensions. Finally, in
Section~\ref{sec:DiscussConclude} we close with a summary and
conclusions. In the appendixes we provide additional information about
some master integrals and, for completeness, results for conversion
factors with complete gauge parameter dependence as well as results for
anomalous dimensions.
\section{Generalities and Notation\label{sec:Generalities}}
In Ref.~\cite{Sturm:2009kb} two regularization invariant
momentum-subtraction schemes with a symmetric subtraction point have
been defined, the so-called $\RISMOM$ and $\RISMOMB$ schemes.  The
$\RISMOM$($\RISMOMB$) scheme can be seen as an extension of the
$\RIp$($\ARI$) scheme from an exceptional to a nonexceptional
subtraction point. The renormalization conditions of these two new
schemes are given in Eqs.~(10), (11), and (A15) of
Ref.~\cite{Sturm:2009kb} and allow for a determination of the
renormalization constants of the fermion field $\Psi$ and the quark mass
$m$ through the computation of the nonsinglet amputated Green's function
$\Lambda_{\hat{O}}$ with the insertion of quark bilinear operators
$\hat{O}$, that is to say, the scalar ($\hat{O}=S$), the pseudoscalar
($\hat{O}=P$), the vector ($\hat{O}=V$), and the axial-vector ($\hat{O}=A$)
operators. Further details can be found in Ref.~\cite{Sturm:2009kb},
whose conventions we follow for these quantities.

In the determination of the light quark mass through lattice simulations,
the $\RISMOM$ and $\RISMOMB$ schemes serve as intermediate schemes
before the conversion of the quark mass to the $\MSbar$ scheme. This
conversion is performed in continuum perturbation theory through the
computation of a matching factor $C^{x}_m$ with the property
\begin{equation}
\label{eq:cmdef}
m^{\msbar}_R=C^{x}_m\*\,m^{x}_R,\qquad x\in\{\rismom, \rismomb\}.
\end{equation}
The index {\scriptsize{$R$}} denotes here and in the following a
renormalized quantity. As shown in Ref.~\cite{Sturm:2009kb} these
conversion factors can be extracted from the amputated Green's function
of the pseudoscalar(or scalar) operator through the equation
\begin{equation}
\label{eq:CP}
(C^{x}_m)^{-1}=C^{x}_q\lim_{m_R\to0}\left.{1\over12i}\Tr\left[\Lambda^{\msbar}_{P,R}(p_1,p_2)\gamma_5\right]\right|_{sym},
\qquad x\in\{\rismom, \rismomb\},
\end{equation}
where $C^{x}_q$ is the corresponding matching factor for the fermion
field which converts the field from the $\RISMOM$($\RISMOMB$) scheme to
the $\MSbar$ scheme via the equation
\begin{equation}
\label{eq:cqdef}
\Psi^{\msbar}_R=\sqrt{C^{x}_q}\*\,\Psi^{x}_R,\qquad x\in\{\rismom, \rismomb\}.
\end{equation}
The matching factor $C^{x}_q$ can be determined through the computation
of the amputated Green's function with the insertion of the
axial-vector(or vector) operator
\begin{equation}
\label{eq:Cqrismom}
(C^{\rismom}_q)^{-1}=\lim_{m_R\to0}\left.{1\over12q^2}\*
\Tr\left[q_{\mu}\Lambda^{\mu,\msbar}_{A,R}(p_1,p_2)\gamma_5\Fslash{q}\right]\right|_{sym}
\end{equation}
and
\begin{equation}
\label{eq:Cqrismomb}
(C^{\rismomb}_q)^{-1}=\lim_{m_R\to0}\left.{1\over48}\Tr\left[
\Lambda^{\mu,\msbar}_{A,R}(p_1,p_2)\*\gamma_5\*\gamma_{\mu}
\right]\right|_{sym},
\end{equation}
where Eqs.~(\ref{eq:Cqrismom}) and (\ref{eq:Cqrismomb}) are distinguished
through the use of two different projectors in the trace with which the
amputated Green's functions are multiplied. The subscript
{\scriptsize{$sym$}} in Eqs.~(\ref{eq:CP}), (\ref{eq:Cqrismom}) and
(\ref{eq:Cqrismomb}) stands for the restriction  of the amputated
Green's function to the symmetric momentum configuration 
\begin{equation}
\label{eq:sym}
\left.f(p_1^2,p_2^2,q^2)\right|_{sym}=\left.f(p_1^2,p_2^2,q^2)\right|_{p_1^2=p_2^2=q^2=-\mu^2},
\end{equation}
where $p_1$ and $p_2$ are the momenta of the external fermions and
$q=p_1-p_2$ is the momentum transfer leaving the operator. The momenta
are also defined in the two-loop diagrams of Fig.~\ref{fig:FeynDia}.
The symbol $\mu^2>0$ denotes the renormalization scale, which we choose
for the $\RISMOM$($\RISMOMB$) scheme to be equal to the one in the
$\MSbar$ scheme.

Also, the tensor operator ($\hat{O}=T$) has interesting applications in
the context of lattice simulations; see
e.g.~Ref.~\cite{Donnellan:2007xr}.  Following the notation and Eq.~(12)
of Ref.~\cite{Sturm:2009kb} one obtains the corresponding matching
factor $C^{x}_T$, which converts the tensor operator from the $x$ scheme 
to the $\MSbar$ scheme,
\begin{equation}
\label{eq:CTdef}
\hat{O}_{R}^{\msbar}=C^{x}_{\hat{O}}\hat{O}_{R}^{x},
\end{equation}
by computing
\begin{equation}
\label{eq:CT}
C^{x}_T=C^{x}_q\lim_{m_R\to0}\left.{1\over144}\Tr\left[\Lambda^{\mu\nu,\,\msbar}_{T,R}(p_1,p_2)\sigma_{\mu\nu}\right]\right|_{sym},\qquad x\in\{\rismom, \rismomb\},
\end{equation}
with $\sigma_{\mu\nu}={i\over2}[\gamma_{\mu},\gamma_{\nu}]$.

The conversion factors in Eqs.~(\ref{eq:cmdef}), (\ref{eq:cqdef}), and
(\ref{eq:CTdef}) are gauge dependent. In the case of the mass conversion
factor this gauge dependence compensates for the one in the mass
renormalization constant which is obtained through lattice simulations;
the NPR procedure of Ref.~\cite{Martinelli:1994ty} is gauge
dependent. The lattice simulations are typically performed in the Landau
gauge ($\xi=0$). For generality we keep the complete dependence on the
gauge parameter and use for the gluon propagator
\begin{equation}
\label{eq:GluonPropagator}
{i\delta^{ab}\over q^2+i\*\ep}\*\left(-g^{\mu\nu}+(1-\xi)\*{q^{\mu}\*q^{\nu}\over q^2+i\*\ep}\right).
\end{equation}
\section{Calculation\label{sec:Calculation}}
The calculation of the two-loop QCD corrections proceeds in two
steps. In the first step, after the generation of the diagrams, all
loop integrals that appear are mapped on a small set of master
integrals.  In the second step these master integrals need to be solved.
\subsection{Manipulation of loop integrals\label{sec:}}
\begin{figure}[t]
\begin{minipage}{2.5cm}
\includegraphics[bb=134 544 278 665,width=2.5cm]{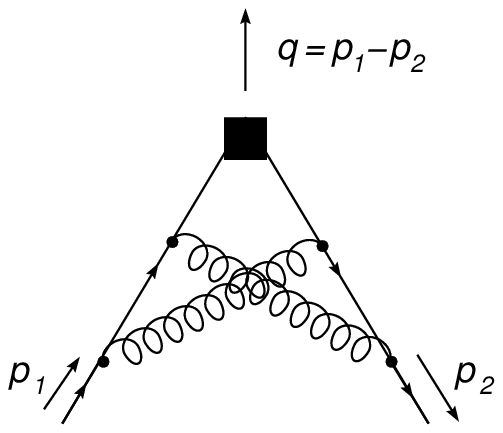}
\begin{center}
\end{center}
\end{minipage}
\begin{minipage}{2.5cm}
\includegraphics[bb=134 544 278 665,width=2.5cm]{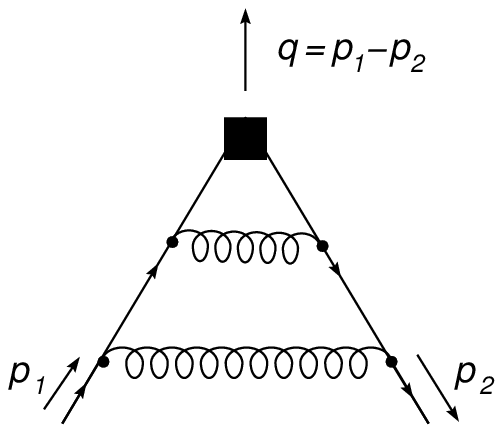}
\begin{center}
\end{center}
\end{minipage}
\begin{minipage}{2.5cm}
\includegraphics[bb=134 544 278 665,width=2.5cm]{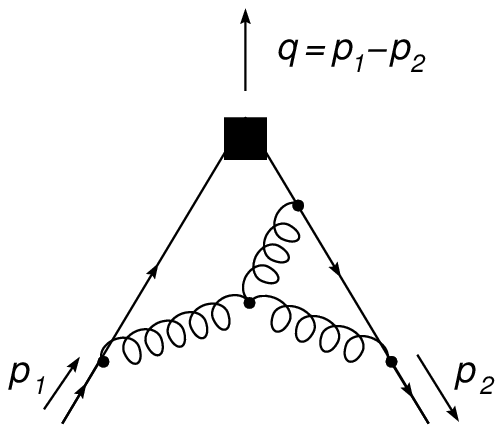}
\begin{center}
\end{center}
\end{minipage}
\begin{minipage}{2.5cm}
\includegraphics[bb=134 544 278 665,width=2.5cm]{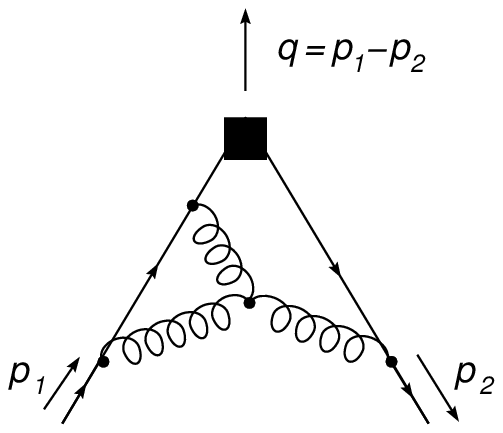}
\begin{center}
\end{center}
\end{minipage}
\begin{minipage}{2.5cm}
\includegraphics[bb=134 544 278 665,width=2.5cm]{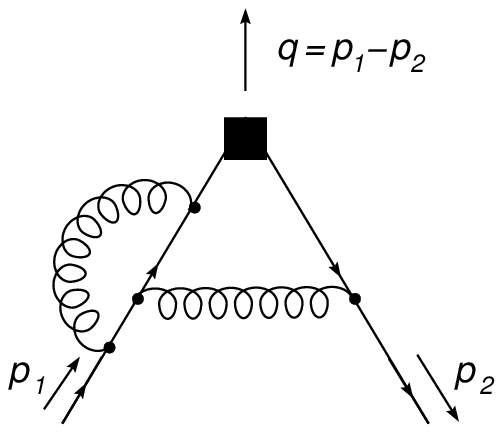}
\begin{center}
\end{center}
\end{minipage}
\begin{minipage}{2.5cm}
\includegraphics[bb=134 544 278 665,width=2.5cm]{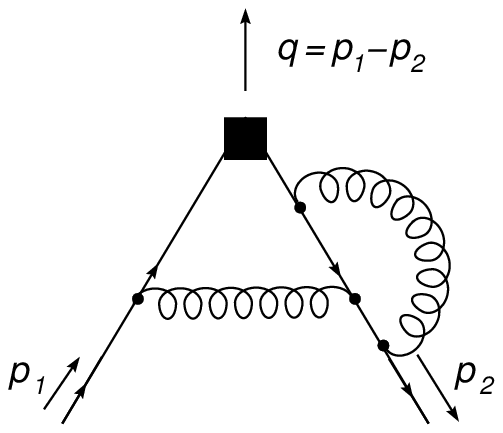}
\begin{center}
\end{center}
\end{minipage}\\[0.55cm]
\mbox{}\hspace{1.3cm}
\begin{minipage}{2.5cm}
\includegraphics[bb=134 544 278 665,width=2.5cm]{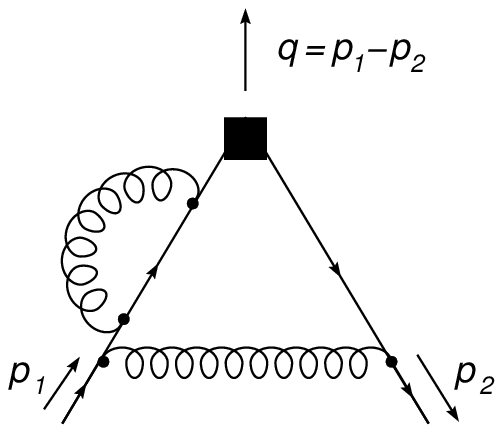}
\begin{center}
\end{center}
\end{minipage}
\begin{minipage}{2.5cm}
\includegraphics[bb=134 544 278 665,width=2.5cm]{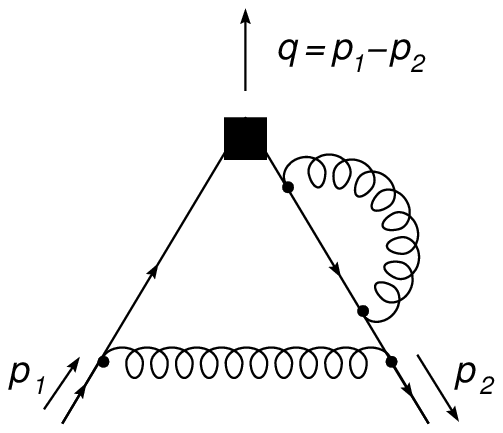}
\begin{center}
\end{center}
\end{minipage}
\begin{minipage}{2.5cm}
\includegraphics[bb=134 544 278 665,width=2.5cm]{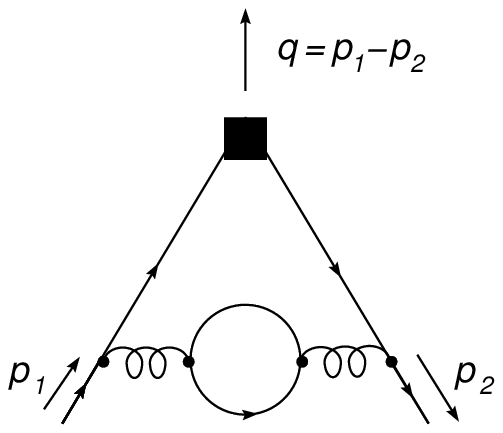}
\begin{center}
\end{center}
\end{minipage}
\begin{minipage}{2.5cm}
\includegraphics[bb=134 544 278 665,width=2.5cm]{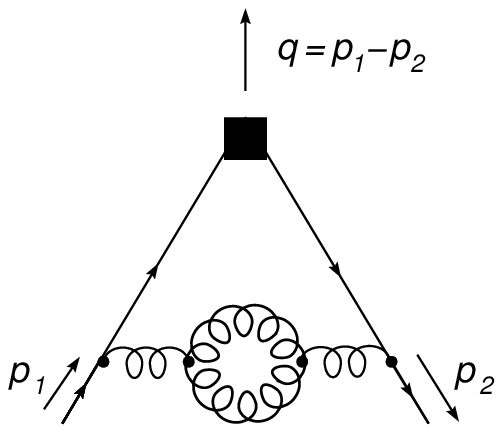}
\begin{center}
\end{center}
\end{minipage}
\begin{minipage}{2.5cm}
\includegraphics[bb=134 544 278 665,width=2.5cm]{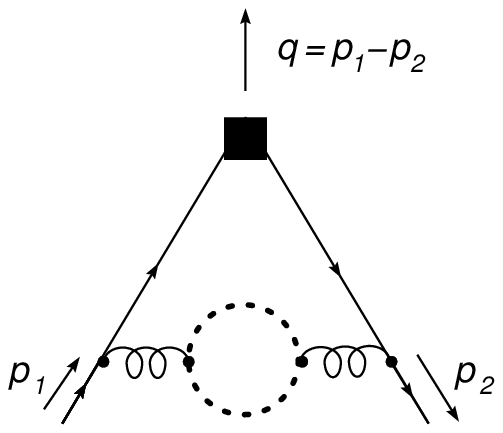}
\begin{center}
\end{center}
\end{minipage}
\caption{\label{fig:FeynDia}Two-loop Feynman diagrams which contribute
  to the computation of the nonsinglet, amputated Green's
  functions. The spiral lines denote gluons, the solid lines represent
  fermions, the dashed lines are ghost fields, and the black boxes indicate the
  inserted operator. }
\end{figure}

In order to generate the required Feynman diagrams, shown in
Fig.~\ref{fig:FeynDia}, we have used the program
{\tt{QGRAF}}~\cite{Nogueira:1991ex}. We identify the different
topologies and adopt the proper notation with the help of the packages
{\tt{q2e}} and
{\tt{exp}}\cite{Seidensticker:1999bb,Seidensticker:1999,Harlander:1997zb}. This
allows us to prepare an output which can straightforwardly be used to
perform the reduction to master integrals. The reduction is achieved
with the traditional integration-by-parts (IBP) method~\cite{Chetyrkin:1981qh}
in combination with Laporta's
algorithm~\cite{Laporta:1996mq,Laporta:2001dd} and has been carried out
with a
{\tt{FORM}}~\cite{Vermaseren:2000nd,Vermaseren:2002rp,Tentyukov:2006ys}
based implementation. The rational functions in the space-time dimension
$d$, which appear while solving the arising linear system of equations,
have been simplified with the program {\tt{FERMAT}}~\cite{Lewis}.
In the case of the computation of the amputated Green's function with
the insertion of the pseudoscalar or axial-vector operator, we use a
naive anticommuting definition of $\gamma_5$ for the treatment of
$\gamma_5$ in dimensional
regularization~\cite{tHooft:1972fi,Breitenlohner:1977hr}, which is a
self-consistent prescription for the flavor nonsinglet contributions
considered here~\cite{Trueman:1979en,Larin:1993tq}.

\subsection{Master integrals}
\begin{figure}[!h]
\begin{center}
\begin{minipage}[t]{2.3cm}
  \begin{center}
\includegraphics[bb=175 535 283 590,width=2.3cm]{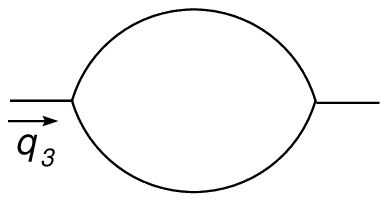}\\
    $\M{1}{2}{1}$
  \end{center}
\end{minipage}
\begin{minipage}[t]{2.3cm}
  \begin{center}
\includegraphics[bb=176 489 353 616,width=2.3cm]{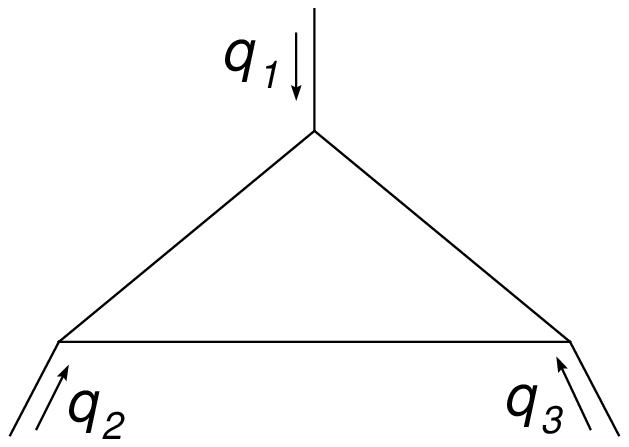}\\
    $\M{1}{3}{1}$
  \end{center}
\end{minipage}
\hspace{1cm}
\begin{minipage}[t]{2.3cm}
  \begin{center}
\includegraphics[bb=175 535 353 590,width=2.3cm]{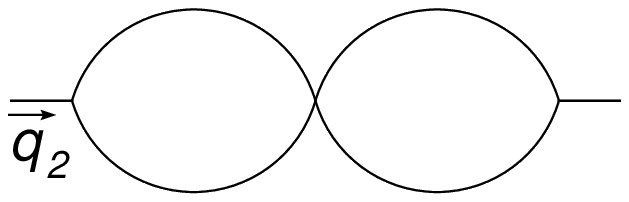}\\
    $\M{2}{4}{1}$
  \end{center}
\end{minipage}
\hspace{0.2cm}
\begin{minipage}[t]{2.3cm}
  \begin{center}
\includegraphics[bb=176 490 353 695,width=2.3cm]{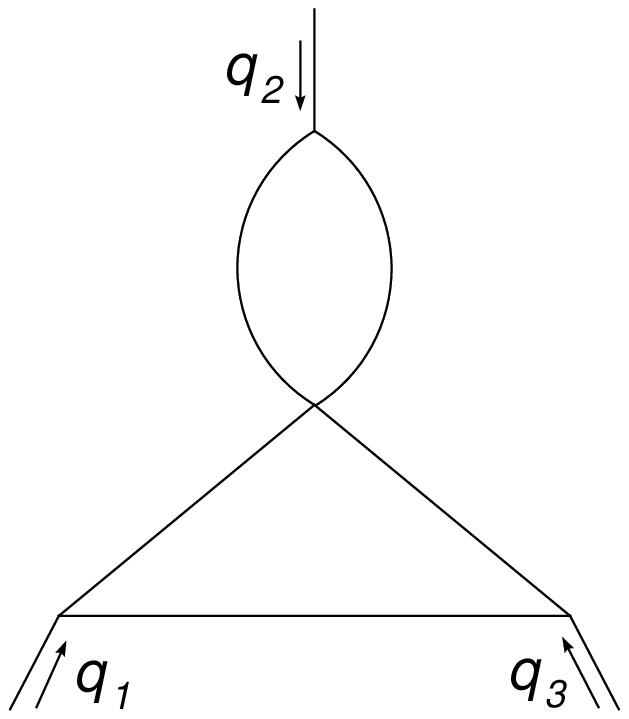}\\
    $\M{2}{5}{1}$
  \end{center}
\end{minipage}\\[0.2cm]
\hspace{0.2cm}
\begin{minipage}[t]{2.3cm}
  \begin{center}
\includegraphics[bb=176 509 353 616,width=2.3cm]{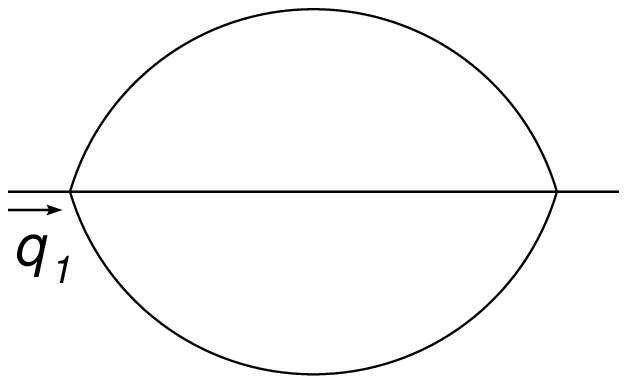}
    $\M{2}{3}{1}$
  \end{center}
\end{minipage}
\hspace{0.2cm}
\begin{minipage}[t]{2.3cm}
  \begin{center}
\includegraphics[bb=176 491 353 695,width=2.3cm]{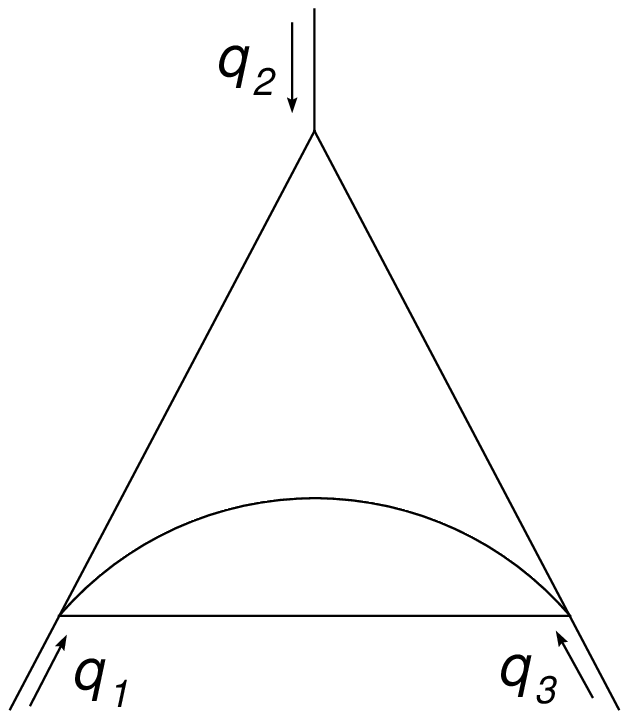}
    $\M{2}{4}{2}$
  \end{center}
\end{minipage}
\hspace{0.2cm}
\begin{minipage}[t]{2.3cm}
  \begin{center}
\includegraphics[bb=176 492 353 695,width=2.3cm]{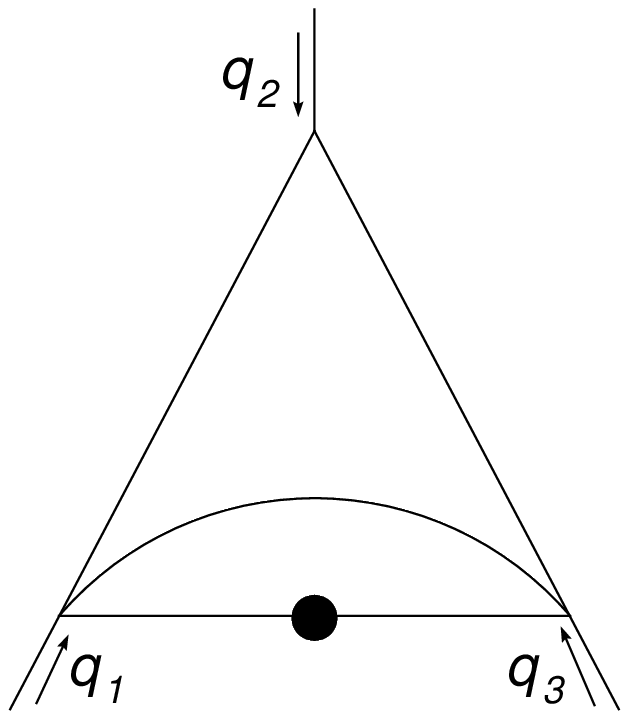}
    $\M{2}{4}{3}$
  \end{center}
\end{minipage}
\hspace{0.2cm}
\begin{minipage}[t]{2.3cm}
  \begin{center}
\includegraphics[bb=176 492 353 695,width=2.3cm]{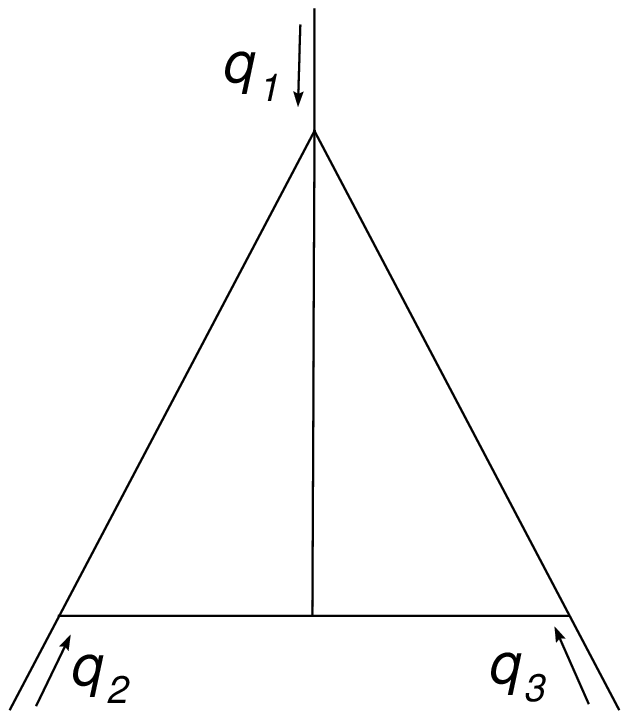}
    $\M{2}{5}{2}$
  \end{center}
\end{minipage}
\hspace{0.2cm}
\begin{minipage}[t]{2.3cm}
  \begin{center}
\includegraphics[bb=176 492 353 695,width=2.3cm]{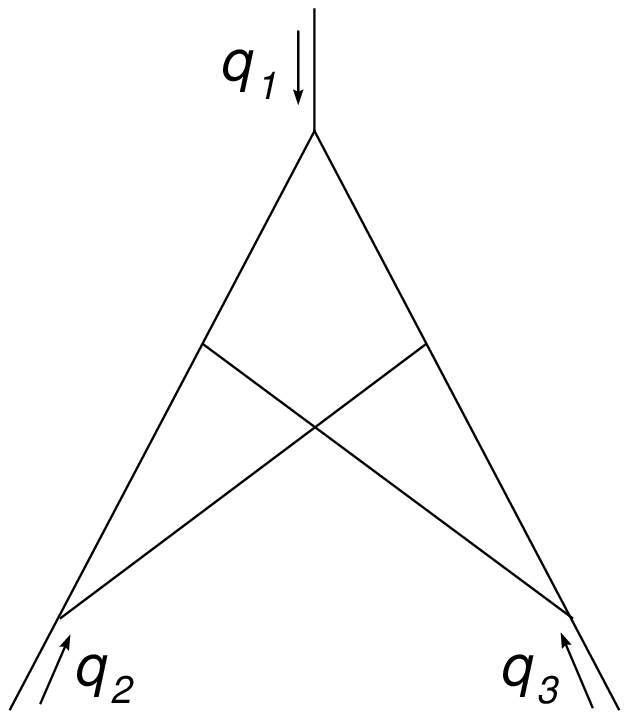}
    $\M{2}{6}{1}$
  \end{center}
\end{minipage}
\end{center}
%%%%
\caption{\label{fig:MI}There appear two one-loop master integrals and
  seven two-loop master integrals from which two are factorized.
The symbol $\M{l}{p}{r}$ denotes a $l$-loop topology
with $p$ lines. The number $r$ serves as a running number to enumerate
topologies with the same number of loops and propagators.}
\end{figure}

After the IBP reduction of the two-loop amplitude, seven
massless master integrals survive. The one- and two-loop master integrals are
shown in Fig.~\ref{fig:MI}.
They are defined in Minkowskian space in $d=4-2\*\vep$ space-time
dimensions by
\begin{equation}
\begin{array}{l@{\quad}l}
\nonumber
%%%
\M{1}{2}{1}=\mu^{2\*\vep} e^{\vep\*\gamma_E}\!\!\displaystyle\int\!
\frac{d^d\ell_1}{i\pi^{d/2}}
\frac{1}{D_1 D_2}, &
%%%
\M{1}{3}{1}=\mu^{2\*\vep} e^{\vep\*\gamma_E}\!\!\displaystyle\int\!
\frac{d^d\ell_1}{i\pi^{d/2}}
\frac{1}{D_1 D_2 D_3},
\end{array}
\end{equation}
%%%%%%%%%%%%%%
\begin{equation}
\begin{array}{l@{\quad}l}
\nonumber
%%%
\M{2}{3}{1}=\mu^{4\*\vep} e^{2\*\vep\*\gamma_E}\!\!\displaystyle\int\!
\frac{d^d\ell_1}{i\pi^{d/2}}\frac{d^d\ell_2}{i\pi^{d/2}}
\frac{1}{D_1 D_4 D_7}, &
%%%
\!\!\!\!\!\!\!\M{2}{4}{1}=\mu^{4\*\vep} e^{2\*\vep\*\gamma_E}\!\!\displaystyle\int\!
\frac{d^d\ell_1}{i\pi^{d/2}}\frac{d^d\ell_2}{i\pi^{d/2}}
\frac{1}{D_1 D_3 D_4 D_5},\\
%%%
\M{2}{4}{2}=\mu^{4\*\vep} e^{2\*\vep\*\gamma_E}\!\!\displaystyle\int\!
\frac{d^d\ell_1}{i\pi^{d/2}}\frac{d^d\ell_2}{i\pi^{d/2}}
\frac{1}{D_1 D_4 D_5 D_7}, &
%%%
\!\!\!\!\!\!\!\M{2}{4}{3}=\mu^{4\*\vep} e^{2\*\vep\*\gamma_E}\!\!\displaystyle\int\!
\frac{d^d\ell_1}{i\pi^{d/2}}\frac{d^d\ell_2}{i\pi^{d/2}}
\frac{1}{D_1^2 D_4 D_5 D_7},\\
%%%
\M{2}{5}{1}=\mu^{4\*\vep} e^{2\*\vep\*\gamma_E}\!\!\displaystyle\int\!
\frac{d^d\ell_1}{i\pi^{d/2}}\frac{d^d\ell_2}{i\pi^{d/2}}
\frac{1}{D_1 D_2 D_3 D_4 D_5},&
%%%
\!\!\!\!\!\!\!\M{2}{5}{2}=\mu^{4\*\vep} e^{2\*\vep\*\gamma_E}\!\!\displaystyle\int\!
\frac{d^d\ell_1}{i\pi^{d/2}}\frac{d^d\ell_2}{i\pi^{d/2}}
\frac{1}{D_1 D_2 D_4 D_5 D_6 },\\
%%%
\M{2}{6}{1}=\mu^{4\*\vep} e^{2\*\vep\*\gamma_E}\!\!\displaystyle\int\!
\frac{d^d\ell_1}{i\pi^{d/2}}\frac{d^d\ell_2}{i\pi^{d/2}}
\frac{1}{D_1 D_2 D_4 D_5 D_6 D_7},
%%%
\end{array}
\end{equation}
%%%%%%
\begin{equation}
\nonumber
\end{equation}
with the denominators
\begin{equation}
\begin{array}{l@{\quad}l@{\qquad}l}
\nonumber
D_1=\ell_1^2 + i\*\ep, &         
D_2=(\ell_1 + q_3)^2 + i\*\ep, &
D_3=(\ell_1 - q_2)^2 + i\*\ep, \\
D_4=\ell_2^2 + i\*\ep, &	        
D_5=(\ell_2 - q_2)^2 + i\*\ep, &
D_6=(\ell_1 - \ell_2)^2 + i\*\ep,\\
D_7=(\ell_1- \ell_2 - q_1)^2 + i\*\ep,
\end{array}
\end{equation}
where $\ell_1$ and $\ell_2$ are loop momenta and $q_1$, $q_2$, and $q_3$
are external momenta with $q_1^2=q_2^2=q_3^2=-\mu^2$.
The symbol $e\simeq2.71828$ is Euler's number and
$\gamma_E\simeq0.577216$ is the Euler-Mascheroni constant.   

The one- and two-loop self-energy integrals $\M{1}{2}{1}$, $\M{2}{3}{1}$
are well known to all orders in $\vep$. The factorized master integrals
$\M{2}{4}{1}$ and $\M{2}{5}{1}$ can be obtained from taking the product
of the two one-loop integrals. Each of the master integrals
$\M{2}{4}{2}$ and $\M{2}{4}{3}$ can be written as a product of a scalar
one-loop two-point function and a one-loop three-point function
$\M{1}{3}{1}$, which has a noninteger power of one of the propagators.
The latter has been determined analytically in terms of hypergeometric
functions to all orders in $\vep$ in
Ref.~\cite{Davydychev:1992xr}. Algorithms to expand generalized
hypergeometric functions in a small parameter, like $\vep$, have been
developed in Refs.~\cite{Moch:2001zr,Moch:2005uc}. For the practical
implementation we use the {\tt{Mathematica}} packages {\tt{HypExp}} and
{\tt{HPL}}~\cite{Maitre:2005uu,Maitre:2007kp,Huber:2007dx}. Expansions
of hypergeometric functions can also be found in
Refs.~\cite{Davydychev:2003mv,Kalmykov:2006pu}. The master integrals
$\M{2}{5}{2}$ and $\M{2}{6}{1}$ can be taken from
Ref.~\cite{Usyukina:1994iw} in the special case of symmetric external
momenta $q_i^2=-\mu^2$ for $i=1,2,3$. Results for the master integrals
in terms of harmonic polylogarithms can also be found in
Ref.~\cite{Birthwright:2004kk}.

The results expanded in $\vep$ to the maximally required order read
\begin{eqnarray}
\label{eq:M121}
\M{1}{2}{1}&=&
   {1\over\vep}
 + 2
 + \vep\*\left[
   4
 - {\pi^2\over12}
        \right]
 + \vep^2\*\left[
   8 
 - {\pi^2\over6} 
 - {7\over3}\*\z3
        \right]
 +\Order{\vep^3},\\
%%%%%%%%%%%%%%%%
\label{eq:M131}
\mu^2\M{1}{3}{1}&=&\!\!\!\!
   \left({2\over3}\*\pi\right)^2\!\!
 - {2\over3}\p{3}
 +\vep\*\left[
   12\*\s{3}{6}
 - {35\over108}{\pi^3\over\sqrt{3}}
 - {\Log{3}{2}\*\pi\over4\sqrt{3}}
       \right]
 +\vep^2\*\mathcal{H}_{31}^{(2)}
%= - 6.114776...
%
\!\! +\!\Order{\vep^3}\!,\\
%%%%%%%%%%%%%%%%
\label{eq:M241}
\M{2}{4}{1}&=& 
   {1\over\vep^2}
 + {4\over\vep}
 + 12
 - {\pi^2\over6}
 +\vep\*\left[
   32 
 - {2\over3}\*\pi^2 
 - {14\over3}\*\z3
       \right]
+\Order{\vep^2},\\
%%%%%%%%%%%%%%%%
\label{eq:M251}
\mu^2\M{2}{5}{1}&=&
 {1\over \vep}\*\left[
   \left({2\over3}\*\pi\right)^2
 - {2\over 3}\*\p{3} 
               \right]
 + 12\*\s{3}{6}
 - {35\over108}\*{\pi^3\over\sqrt{3}} 
 - {\Log{3}{2}\*\pi \over 4\*\sqrt{3}}
 - {4\over3}\*\p{3}
 \nonumber\\&&
 + {8\over9}\*\pi^2
 + \vep\left[
   24\*\s{3}{6} 
 - {35\*\pi^3\over54\*\sqrt{3}}
 - {\Log{3}{2}\*\pi\over2\*\sqrt{3}} 
 - {8\over3}\*\p3 
 + {16\over9}\*\pi^2 
 + {\pi^2\over18} \*\p3
 \right.
 \nonumber\\&&
 \left.
 - {\pi^4\over27} 
 + \mathcal{H}_{31}^{(2)}
       \right]
% = - 21.63777...
 +\Order{\vep^2},\\
%%%%%%%%%%%%%%%%
\label{eq:M231}
{1\over\mu^2}\M{2}{3}{1}&=& 
 {1\over4\*\vep} 
 + {13\over8} 
 + \vep\*\left[
   {115\over16} 
 - {\pi^2\over24}
       \right]
 + \vep^2\*\left[
   {865\over32} 
 - {13\over48}\*\pi^2 
 - {8\over3}\*\z3
         \right]
 + \Order{\vep^3},\\
%%%%%%%%%%%%%%%%
\label{eq:M242}
\M{2}{4}{2}&=&
 {1\over2\*\vep^2}
 + {5\over2\*\vep}
 + {19\over2} 
 - {1\over3}\*\p3 
 + {5\over36}\*\pi^2
 +\vep\*\left[
   {65\over2} 
 - 12\*\s{2}{2} 
 + 6\*\s{2}{6} 
\right.\nonumber\\&&\left.
 + 8\*\s{3}{2}
 - 4\*\s{3}{6} 
 - {67\over324\*\sqrt{3}}\*\pi^3 
 - {\pi\*\nlog{3}\over2\*\sqrt{3}} 
 - {\Log{3}{2}\*\pi\over12\*\sqrt{3}}
 - {2\over3}\*\p3
\right.\nonumber\\&&\left.
 + {\pi^2\over36}
 - {10\over3}\*\z3
%=16.50225...
 \right]
 + \Order{\vep^2},\\
%%%%%%%%%%%%%%%%
\label{eq:M243}
\mu^2\M{2}{4}{3}&=&
 {1\over\vep}\left(
   {2\over3}\*\p3 
 - \left({2\over3}\*\pi\right)^2
            \right)  
 + {\pi\*\Log{3}{2}\over4\*\sqrt{3}} 
 + {35\*\pi^3\over108\*\sqrt{3}} 
 - 12\*\s{3}{6}
\nonumber\\
&+&\vep\*\left[
   {\pi^4\over27}
 - {\pi^2\over18}\*\p3 
 + \mathcal{H}_{43}^{(2)}
       \right]
%=   10.53216...
%
 + \Order{\vep^2},\\
%%%%%%%%%%%%%%%%
\label{eq:M252}
\mu^2\M{2}{5}{2}&=& 
   {2\over27}\*\pi^4 
 - {1\over36}\*\ppp{3} 
 + \Order{\vep},\\
%%%%%%%%%%%%%%%%
\label{eq:M261}
\mu^4\M{2}{6}{1}&=&
 \left[
   \left({2\over3}\*\pi\right)^2
 - {2\over3}\p{3}\right]^2
 + \Order{\vep},
\end{eqnarray}		  
where $\Psi(x)$ is the digamma function
$\Psi(x)=\Gamma'(x)/\Gamma(x)$; here the prime denotes the derivative of
the $\Gamma$ function. We also define
$s_n(x)={1\over\sqrt{3}}\mbox{Im}\left[\mbox{Li}_n\left({e^{ix}\over\sqrt{3}}\right)\right]$
with the polylogarithm function
$\mbox{Li}_n(z)=\sum_{k=1}^{\infty}{z^k\over k^n}$. The symbol
$\zeta_n=\mbox{Li}_n(1)$ is the Riemann zeta function. The constants 
$\mathcal{H}_{31}^{(2)}$ and $\mathcal{H}_{43}^{(2)}$ can be expressed
in terms of harmonic polylogarithms and are discussed in
Appendix~\ref{sec:MIpolylog}.

The higher orders in $\vep$ of some of the master integrals are needed
due to the appearance of so-called spurious poles which arise while
solving the linear system of IBP equations. An alternative approach is
to find a different set of master integrals through the method of
$\vep$-finite basis~\cite{Chetyrkin:2006dh} which exploits the fact that
the choice of the master integrals is not unique and which requires only
the evaluation of the master integrals up to the order $\vep^0$.

The master integrals constitute an essential input of our calculation,
so we have checked the results using traditional Feynman
parametrizations, partially analytically and partially through the
numerical evaluation of the resulting Feynman integrals in order to ensure their
correctness.
\section{Results\label{sec:Results}}
\subsection{Matching factors\label{sec:matchfactors}}
In this section we give the conversion factors up to two-loop order in
perturbative QCD for the fermion field and the mass parameter. This is  
accomplished as previously described in
Section~\ref{sec:Generalities}. For this purpose we decompose the
matching factors into a part which depends on the gauge parameter and 
a part which is gauge parameter free,
\begin{equation}
\label{eq:decomp}
C^{x}_{y}=C^{x}_{y,L}+C^{x}_{y,\xi},
\quad x\in\{\rismom,\;\rismomb\},	
\quad y\in\{q,\;m\},	
\end{equation}
where $C^{x}_{y,\xi}$ contains all the terms which depend on the
gauge parameter $\xi$, whereas $C^{x}_{y,L}$ is the conversion factor in
the Landau gauge. In the following we present the contributions in the
Landau gauge; for completeness, we also provide the gauge dependent
terms in Appendix \ref{sec:Convxi}. The one-loop results have been
determined in Ref.~\cite{Sturm:2009kb}.

Let us start with the conversion factors for the fermion field of
Eqs.~(\ref{eq:Cqrismom}) and (\ref{eq:Cqrismomb}). They are important for
the light quark mass renormalization procedure in Eq.~(\ref{eq:CP}), but
also enter in the renormalization of other interesting quantities,
i.e. any multiquark operator in such schemes, like, for example, the
$B_K$ parameter. With the help of Ward-Takahashi identities it has been
shown in Ref.~\cite{Sturm:2009kb} that $C^{\rismom}_{q}=C^{\rip}_{q}$.
We have checked that our order $\alpha_s^2$ result for $C^{\rismom}_{q}$
is in agreement with $C^{\rip}_{q}$ from
Ref.~\cite{Chetyrkin:1999pq}. The fermion field matching factor in the
$\RISMOMB$ scheme at two-loop order is new. The result is given by
\begin{eqnarray}
\label{eq:CqLrismomb}
C^{\rismomb}_{q,L}&=&
1 + \left({\alpha_s\over4\*\pi}\right)\*\CF 
  + \left({\alpha_s\over4\*\pi}\right)^2\*\left\{
   \CF\*\nf\*\TF \* {5\over 18}
 \right. \nonumber\\&& \left.\hspace{-2cm} 
  + \CF^2\*\left[
- {1\over 8 }
+ {13\*\pi^2\over 3} 
+ {29\*\pi^3\over 162\*\sqrt{3} }
+ {2\*\pi^4\over 81} 
- 2\*\z3 
+ {2\*\pi\*\nlog{3}\over \sqrt{3}}
- {\pi\*\Log{3}{2}\over 6\*\sqrt{3}} 
 \right. \right. \nonumber\\&& \left.\left. 
- 24\*\s{2}{6} 
+ 48\*\s{2}{2} 
+ 40\*\s{3}{6} 
- 32\*\s{3}{2}
- {13\over 2}\*\p{3}  
\right. \right. \nonumber\\&& \left.\left.
- {8\*\pi ^2\over 27}\*\p{3}
+ {2\over 9}\*\p{3}^2
+ {1\over 36 }\*\ppp{3}
        \right] 
\right.\nonumber\\&& \left.\hspace{-2cm}
+ \CA\*\CF\* \left[
- {31\over 72} 
- {41\*\pi^2\over 18} 
- {145\*\pi^3\over 1296\*\sqrt{3}} 
- {7\*\pi^4\over 81 }
- \z3 
- {5\*\pi\*\nlog{3}\over 4\*\sqrt{3}} 
+ {5\*\pi\*\Log{3}{2}\over 48\*\sqrt{3}}  
\right. \right. \nonumber\\&& \left.\left.
+ 15\*\s{2}{6} 
- 30\*\s{2}{2} 
- 25\*\s{3}{6} 
+ 20\*\s{3}{2}
+ {41 \over 12}\*\p{3} 
\right. \right. \nonumber\\&& \left.\left.
+ {4\*\pi^2\over 27}\*\p{3}
- {1\over 9}\*\p{3}^2
+ {1\over 72}\*\ppp{3}
       \right]
\right\}
+\Order{\alpha_s^3}\\\nonumber
%%%%%%%%%%%%%%%%%%%%%%%%%%%%%%%%%%%%%%%%%%%
&\stackrel{N_c=3}{\simeq}&\!\!\!\!\!
1 +\!\left({\alpha_s\over4\pi}\right) 1.333333333 
  +\!\left({\alpha_s\over4\pi}\right)^2\!\!\left[9.59901080 + 0.1851851852\nf\right]
  + \Order{\alpha_s^3}.
\end{eqnarray}
The symbols $\CF$ and $\CA$ denote the Casimir operators of the
fundamental and adjoint representations of SU($N_c$). The normalization
of the trace in the fundamental representation of SU($N_c$) is given by
$\TF=1/2$.  For the number of colors $N_c=3$ one obtains $\CF=4/3$ and
$\CA=3$. The number of active fermions is given by $\nf$. 

Starting from two-loop order the strong coupling constant
$\alpha_s$ also has to be renormalized, and it has to be ensured that the
Slavnov-Taylor identities, which provide relations among
the different renormalization constants, are preserved. These relations guarantee that
the renormalized coupling constant remains the same for the different
interaction vertices of the QCD Lagrangian. Therefore, we perform the
renormalization of $\alpha_s$ in the $\MSbar$ scheme with
$\alpha_{s,R}=Z_g^{-2}\*\mu^{-2\*\vep}\*\alpha_{s,B}$. When not
considering the Landau gauge, we also have to perform a renormalization
of the gauge parameter $\xi_R=Z^{-1}_\xi\*\xi_B$ with $Z_{\xi}=Z_{3}$,
where the renormalization constant $Z_3$ is the one of the gluon field
$G^{a,\mu}$ with $G^{a,\mu}_R=1/\sqrt{Z_3}\*G^{a,\mu}_B$. We renormalize
these quantities in the $\MSbar$ scheme (see also
Appendix~\ref{sec:Convxi}).  In general, the ghost field $c^{a}$ also 
needs to be renormalized, $c^{a}_R=1/\sqrt{Z^{c}_3}\*c^{a}_B$, which can
also be done in the $\MSbar$ scheme, but is not needed within this
calculation.

In complete analogy to Eq.~(\ref{eq:CqLrismomb}), the mass conversion
factor, required for the matching of the light quark masses determined
in the $\RISMOM$ scheme to the $\MSbar$ scheme, reads up to two-loop
order in the Landau gauge ($\xi=0$)
\begin{eqnarray}
\label{eq:CmLrismom}
C^{\rismom}_{m,L}=
1&+&\left({\alpha_s\over4\*\pi}\right)\*
  \CF\*\left(
    \p{3}
 - {2\over3}\*\pi^2
 - 4
     \right)
\nonumber\\&& \hspace{-2cm}
 + \left({\alpha_s\over4\*\pi}\right)^2\*\left\{
   \CF\*\TF\*\nf\*\left[ 
   {83\over 6} 
 + {40\*\pi^2\over 27}
 - { 20\over 9 }\*\p{3}
              \right] 
\right.\nonumber\\&&\left.\hspace{-2cm}
 + \CF^2\*\left[
   {19\over 8}
 + {52\*\pi^2\over 9 }
 + {29\*\pi^3\over 162\*\sqrt{3}} 
 + {76\*\pi^4\over 81} 
 + 3\*\Sigma 
 + 4\*\z3 
 + {2\*\pi\*\nlog{3}\over \sqrt{3}} 
 \right. \right. \nonumber\\&&\left. \left.
 - {26 \over 3}\*\p{3} 
 - {52\*\pi^2 \over 27}\*\p{3} 
 + {13 \over 9}\*\p{3}^2 
 - {1 \over 9}\*\ppp{3} 
\right. \right. \nonumber \\&&\left. \left. 
 - {\pi\*\Log{3}{2}\over 6\*\sqrt{3}}
 - 24\*\s{2}{6} 
 + 48\*\s{2}{2} 
 + 40\*\s{3}{6} 
 - 32\*\s{3}{2} 
         \right] 
\right. \nonumber\\[2mm] &&\left.\hspace{-2cm} 
 + \CA\*\CF\*\left[
   {7\*\pi^4\over 81 }
 - {1285\over 24} 
 - {457\*\pi^2\over 54} 
 - {29\*\pi^3\over 324\*\sqrt{3}} 
 + {5\over 2 }\*\Sigma
 + 10\*\z3 
 - {\pi\*\nlog{3} \over \sqrt{3} } 
\right. \right. \nonumber\\[2mm]&& \left.\left.
 - {5\over 72 }\*\ppp{3}
 + 12\*\s{2}{6} 
 - 24\*\s{2}{2} 
 - 20\*\s{3}{6} 
 + 16\*\s{3}{2}  
\right. \right. \nonumber\\[2mm]&& \left. \left. 
 + {\pi\*\Log{3}{2}\over 12\*\sqrt{3} }
 + {457\over 36}\*\p{3} 
 + {8\*\pi^2\over 27}\*\p{3} 
 - {2\over 9 }\*\p{3}^2
            \right] 
 \right\}
 +\Order{\alpha_s^3} \\\nonumber
%%%%%%%%%%%%%
&\stackrel{N_c=3}{\simeq}&\!\!\!\!\!
1 -\!\left({\alpha_s\over4\pi}\right) 0.6455188560
  -\!\left({\alpha_s\over4\pi}\right)^2\!\!\left[22.60768757 - 4.013539470\*\nf\right] 
  + \Order{\alpha_s^3},
\end{eqnarray}
where the symbol $\Sigma$ is given in Appendix~\ref{sec:MIpolylog}.
Similarly, we obtain, for the mass conversion factor up to two-loop order
in the $\RISMOMB$ scheme,
\begin{eqnarray}
\label{eq:CmLrismomb}
C^{\rismomb}_{m,L}&=&
1 +\left({\alpha_s\over4\*\pi}\right)\*
  \CF\*\left(
   \p{3}
 - {2\over3}\*\pi^2
 - 5
      \right)
\nonumber\\&& 
 +\left({\alpha_s\over4\*\pi}\right)^2\*\left\{
 \nf\*\CF\*\TF\*\left[
   {307\over 18 }
 + {40\*\pi^2\over 27 }
 - {20\over 9}\*\p{3}
              \right]
\right.\nonumber\\&& \left. \hspace{-10mm}
 +\CF^2\*\left[
   {65\over 8} 
 + {19\*\pi^2\over 9} 
 + {74\*\pi^4\over 81} 
 + 3\*\Sigma 
 + 6\*\z3 
 - {19\over 6}\*\p{3} 
 - {44\*\pi^2 \over 27}\*\p{3}
 + {11 \over 9}\*\p{3}^2
\right. \right. \nonumber\\&&\left.\left.
 - {5 \over 36}\*\ppp{3}
       \right]
\right.\nonumber\\&&\left.\hspace{-10mm}
 +\CF\*\CA\*\left[
   {14\*\pi^4\over 81} 
 - {2281\over 36}
 - {167\*\pi^2\over 27} 
 + {29\*\pi^3\over 1296\*\sqrt{3} }
 + {5\over 2 }\*\Sigma 
 + 14\*\z3 
 + {\pi\*\nlog{3}\over 4\*\sqrt{3}} 
\right. \right. \nonumber\\&&\left.\left.
 - {\pi\*\Log{3}{2}\over 48\*\sqrt{3} }
 + {167\over 18}\*\p{3} 
 + {4\*\pi^2\over 27}\*\p{3}
 - {1\over 9}\*\p{3}^2
 - 3\*\s{2}{6} 
 + 6\*\s{2}{2} 
\right. \right. \nonumber\\&&\left.\left.
- {1\over 12}\*\ppp{3}
+ 5\*\s{3}{6} 
- 4\*\s{3}{2}
        \right]
\right\}
+\Order{\alpha_s^3}\\\nonumber
%%%%%%%%%%%%%%%%%%%%%%%%%%%%%
&\stackrel{N_c=3}{\simeq}&\!\!\!\!\!
1 -\!\left({\alpha_s\over4\pi}\right) 1.978852189
  -\!\left({\alpha_s\over4\pi}\right)^2\!\!\left[55.03243483 - 6.161687618\*\nf\right] 
  + \Order{\alpha_s^3}.
\end{eqnarray}
We want to mention that the matching factors in Eqs.~(\ref{eq:CmLrismom})
and (\ref{eq:CmLrismomb}) are related to conversion factors of quark
bilinear operators. As shown in Ref.~\cite{Sturm:2009kb} Ward-Takahashi
identities allow us to write the matching factor $C^{x}_P$, which converts
the pseudoscalar operator renormalized in the $x$ scheme to the $\MSbar$
scheme, as the inverse of $C^{x}_m$ for $x\in\{\rismom$, $\rismomb\}$.
Furthermore, the latter are also related to the corresponding matching
factor of the scalar operator $C^{x}_S=C^{x}_P=1/C^{x}_m$,
$x\in\{\rismom$, $\rismomb\}$. For the definition of these matching
factors we adopt the same notations as in Ref.~\cite{Sturm:2009kb}.
For completeness we give here also the result for the matching factor
$C^{x}_{T}$ of the tensor operator and again disentangle $C^{x}_{T}$
into its Landau gauge ($\xi=0$) and its gauge dependent parts in complete
analogy to the decomposition of Eq.~(\ref{eq:decomp}) for $y=T$. The
Landau gauge component is given by
\begin{eqnarray}
\label{eq:CTRISMOM}
C^{\rismom}_{T,L}&=&
1 + \left({\alpha_s\over4\*\pi}\right)\*
   \CF\*\left(
   {1\over 3}\*\p{3} 
 - {2\over 9}\*\pi^2
 - {4\over 3} 
       \right) 
\nonumber\\&& 
 + \left({\alpha_s\over 4\*\pi}\right)^2\*\left\{
   \CF\*\TF\*\nf\*\left[
   {473\over 54} 
 - {20 \over 27 }\*\p{3} 
 + {40 \over 81 }\*\pi^2
                 \right]
\right.\nonumber\\&& \left. \hspace{-10mm}
 + \CF^2\*\left[
   {599\over 24} 
 + {178\over 9 }  \*\p{3} 
 - {4  \over 27  }\*\p{3}^2 
 - {356\over 27}  \*\pi^2 
 + {16 \over 81 } \*\p{3}\*\pi^2 
 + {32 \over 243} \*\pi^4 
\right. \right. \nonumber\\&&\left.\left.
 - {2  \over 27  }\*\ppp{3} 
 - 240\*\s{2}{2} 
 + 120\*\s{2}{6} 
 + 160\*\s{3}{2} 
 - 200\*\s{3}{6} 
\right. \right. \nonumber\\&&\left.\left.
 + \Sigma 
 - {10\*\nlog{3}\*\pi \over \sqrt{3}}
 + {5\*\Log{3}{2}\*\pi\over 6\*\sqrt{3}} 
 - {145\*\pi^3   \over 162\*\sqrt{3}}
 - 12\*\z3
          \right] 
\right.\nonumber\\&&\left.\hspace{-10mm}
 + \CA\*\CF\*\left[
 - {8491\over 216} 
 - {755\over 108}\*\p{3} 
 + {2\over 27}\*\p{3}^2 
 + {755\over 162}\*\pi^2 
 - {8\over 81}\*\p{3}\*\pi^2 
\right. \right. \nonumber\\&&\left.\left.
 + {35\over 243}\*\pi^4 
 - {1\over 24}\*\ppp{3} 
 + 128\*  \s{2}{2} 
 - 64\*   \s{2}{6} 
 - {256\over 3}\*\s{3}{2} 
 + {320\over 3}\*\s{3}{6} 
\right. \right. \nonumber\\&&\left.\left.
 + {5\over 6}\*\Sigma 
 + {16\*\nlog{3}\*\pi\over 3\*\sqrt{3}} 
 - {4\*\Log{3}{2}\*\pi\over 9\*\sqrt{3}}
 + {116\*\pi^3\over 243\*\sqrt{3}}
 + {50\over 3}\*\z3
         \right]
          \right\}
+\Order{\alpha_s^3}\\\nonumber
%%%%%%%%%%%%%%%%%%%%%%%%%%%%%
&\stackrel{N_c=3}{\simeq}&\!\!\!\!\!
1 -\!\left({\alpha_s\over4\pi}\right)\*0.21517295
  -\!\left({\alpha_s\over4\pi}\right)^2\!\!\left[ 43.38395007 - 4.10327859\*\nf\right] 
  + \Order{\alpha_s^3}
\end{eqnarray}
%%%%%%%%%%%%%%%%%%%%%%%%%%%%%%%%%%%%%%%%%%%%%%%%%%%%%%%%%%%%%%%%%%%%
and
%%%%%%%%%%%%%%%%%%%%%%%%%%%%%%%%%%%%%%%%%%%%%%%%%%%%%%%%%%%%%%%%%%%%
\begin{eqnarray}
\label{eq:CTRISMOMB}
C^{\rismomb}_{T,L}&=&
  1 
 + \left({\alpha_s\over4\*\pi}\right)\*
   \CF\*\left(
   {1\over 3}\*\p{3} 
 - {2\over 9}\*\pi^2
 - {1\over 3} 
      \right) 
\nonumber\\&& 
 + \left({\alpha_s\over4\*\pi}\right)^2\*\left\{
   \CF\*\TF\*\nf\*\left[
   {299\over 54} 
 - {20 \over 27 }\*\p{3} 
 + {40 \over 81 }\*\pi^2
                 \right]
\right.\nonumber\\&& \left. \hspace{-10mm}
 + \CF^2\*\left[
   {183\over 8} 
 + {245\over 18}\*\p{3} 
 + {2\over 27}\*\p{3}^2 
 - {245\over 27}\*\pi^2 
 - {8\over 81}\*\p{3}\*\pi^2 
 + {38\over 243}\*\pi^4 
\right. \right. \nonumber\\&&\left.\left.
 - {5\over 108}\*\ppp{3} 
 - 192\*\s{2}{2} 
 + 96\* \s{2}{6} 
 + 128\*\s{3}{2} 
 - 160\*\s{3}{6} 
 + \Sigma 
\right. \right. \nonumber\\&&\left.\left.
 - {8\*\nlog{3}\*\pi\over \sqrt{3}}
 + {2\*\Log{3}{2}\*\pi\over  3\*\sqrt{3}} 
 - {58\*\pi^3\over 81\*\sqrt{3}}
 - 14\*\z3
         \right] 
\right.\nonumber\\&& \left. \hspace{-10mm}
 + \CA\*\CF\*\left[
 - {3185\over 108}
 - {193\over 54}\*\p{3} 
 - {1\over 27}\*\p{3}^2 
 + {193\over 81}\*\pi^2 
 + {4\over 81}\*\p{3}\*\pi^2 
 + {14\over 243}\*\pi^4 
\right. \right. \nonumber\\&&\left.\left.
 - {1\over 36}\*\ppp{3} 
 + 98\*   \s{2}{2} 
 - 49\*   \s{2}{6} 
 - {196\over 3}\*\s{3}{2} 
 + {245\over 3}\*\s{3}{6} 
 + {5\over 6}\*\Sigma 
\right. \right. \nonumber\\&&\left.\left.
 + {49\*\nlog{3}\*\pi\over 12\*\sqrt{3}} 
 - {49\*\Log{3}{2}\*\pi\over 144\*\sqrt{3}} 
 + {1421\*\pi^3\over 3888\*\sqrt{3}}
 + {38\over 3}\*\z3
         \right]
        \right\}
+\Order{\alpha_s^3}\\\nonumber
%%%%%%%%%%%%%%%%%%%%%%%%%%%%%
&\stackrel{N_c=3}{\simeq}&\!\!\!\!\!
1 +\!\left({\alpha_s\over4\pi}\right) 1.11816038
  -\!\left({\alpha_s\over4\pi}\right)^2\!\!\left[ 8.607630493 - 1.955130440\*\nf\right] 
  + \Order{\alpha_s^3}.
\end{eqnarray}
The gauge dependent component can be found in Appendix~\ref{sec:Convxi}.
The one-loop order in the $\RISMOM$ scheme has already been computed in
Ref.~\cite{Sturm:2009kb}. The result in the $\RIp$ scheme is available
up to three-loop order in Ref.~\cite{Gracey:2003yr}.%
\subsection{Comparison with results using exceptional momenta\label{sec:Comparison}}
In order to analyze the quality of the new two-loop order of the mass
conversion factors in Eqs.~(\ref{eq:CmLrismom}) and (\ref{eq:CmLrismomb}),
we compare them to the results of the traditional $\RIp$ and $\ARI$
schemes.  For this purpose we evaluate the matching factors numerically,
again in the Landau gauge, for $\nf=3$ flavors. For the strong coupling
constant we insert $\alpha_s/\pi=0.1$
\footnote{The choice of this value corresponds approximately to the
 value of $\alpha_s/\pi$ at a scale of 2~GeV and is sufficient to
 illustrate the effect of the new results. It allows the reader to
 easily replace it by her/his own value of $\alpha_s$.}.  The results in
 the $\RISMOM$ versus the $\RIp$ scheme for $N_c=3$ read
\begin{eqnarray}
%%%%
\label{eq:ncmrismom}
C^{\rismom}_{m,L}&=&1 - 0.0161380... - 0.00660442...,\\
%%%%
%C^{\rip}_m&=&1  
%-{\alpha_s\over4\pi} 5.33333333...
%-\left({\alpha_s\over4\pi}\right)^2 121.373561...
%-\left({\alpha_s\over4\pi}\right)^3 2933.32893...
%+\Order{\alpha_s^4},\\
%%
\label{eq:ncmrip}
C^{\rip}_{m,L}&=&1 - 0.1333333...- 0.07585848...- 0.0556959...,%\\
\end{eqnarray}
%%%%%
where each term stands for the next order in perturbative QCD, i.e. the
leading, next-to-leading, and next-to-next-to-leading order results.
Similarly follow the results for the $\RISMOMB$ versus the $\ARI$ scheme
%%%%%
\begin{eqnarray}
%%%%%
\label{eq:ncmrismomb}
C^{\rismomb}_{m,L}&=&1
- 0.0494713...
- 0.0228421...,\\
%%%%%
%C^{\ari}_m&=&1 
%-{\alpha_s\over4\pi} 5.33333333...
%-\left({\alpha_s\over4\pi}\right)^2 130.540228...
%-\left({\alpha_s\over4\pi}\right)^3 3857.65686...
%+\Order{\alpha_s^4},\\
%%
\label{eq:ncmari}
C^{\ari}_{m,L}&=&1 - 0.1333333... - 0.0815876... - 0.0602759....%\\
%%%%%
\end{eqnarray}
Both the $\ARI$ and $\RIp$ schemes are known to three-loop order, where
we have taken the results of Ref.~\cite{Chetyrkin:1999pq}.  One can
observe that the matching factors of the schemes with a symmetric
subtraction point in Eqs.~(\ref{eq:ncmrismom}) and (\ref{eq:ncmrismomb})
show a much better convergence behavior than the ones in the $\RIp$ and
$\ARI$ schemes of Eqs.~(\ref{eq:ncmrip}) and (\ref{eq:ncmari}) which are
characterized by an exceptional momentum-subtraction point. This
observation has already been made in Ref.~\cite{Sturm:2009kb} at
one-loop order and is now confirmed at the two-loop level. The size of
the three-loop corrections is about 6\% of the leading order result for
both the $\ARI$ and $\RIp$ schemes, whereas in the $\RISMOM$ scheme
the two-loop result is already significantly smaller, being of about only
7 per mill%
\footnote{Since the percentage correction of the two-loop term in the
$\RISMOM$ scheme is very small, its size is more sensitive to the exact
value of $\alpha_s$ than in the other schemes. Thus a slightly smaller
value of $\alpha_s/\pi$ leads to a slightly smaller correction of about
0.6\% for the $\RISMOM$ scheme.}
of the leading order one. Similarly, for the $\RISMOMB$ scheme the size
of the two-loop term is of about 2\%, which is also smaller than the
three-loop terms of the two schemes with an exceptional subtraction
point.  The use of these new results in light up-, down-, and
strange-quark mass determinations in the context of NPR will allow us to
reduce the uncertainties related to the matching procedure to the
$\MSbar$ scheme, due to a smaller truncation error of the perturbative
series, which will significantly reduce the error of these light quark
masses obtained in this approach.
\subsection{Anomalous dimensions}
The mass anomalous dimension of the $\RISMOM$ and $\RISMOMB$ schemes is
used to run the quark masses to different energy scales. It is
defined by
\begin{equation}
\label{eq:gammam}
\gamma^x_m={d\log{m^x(\mu)}\over d\log{\mu^2}}=
-\gamma^{(0),x}_m\*      \frac{\alpha_s}{4\*\pi}       
-\gamma^{(1),x}_m\*\left(\frac{\alpha_s}{4\*\pi}\right)^2
-\gamma^{(2),x}_m\*\left(\frac{\alpha_s}{4\*\pi}\right)^3
+\mathcal{O}(\alpha_s^4)\,,
\end{equation}
where the superscript $x$ stands, similar to Eq.~(\ref{eq:decomp}),
for the scheme with $x\in\{\rismom$, $\rismomb$, $\msbar\}$.  Both mass
anomalous dimensions are related to the $\MSbar$ mass anomalous
dimension $\gamma_m^{\msbar}$ through the conversion factors $C^{x}_m$
presented in Section~\ref{sec:matchfactors} (see
Ref.~\cite{Chetyrkin:1999pq}). In the Landau gauge holds
\begin{equation}
\gamma_m^{x}=\gamma_m^{\msbar}-\beta\*
 {\partial \log{C_{m,L}^x}\over
  \partial\tfrac{\alpha_s}{4\pi}},
\quad x\in\{\rismom,\;\rismomb\},
\end{equation}
where the QCD $\beta$ function and $\gamma_m^{\msbar}$ are given, for
completeness, in Appendix~\ref{sec:MSbarAnomalous}.  The one-loop mass
anomalous dimensions in the $\RISMOM$ and $\RISMOMB$ schemes are equal to
the one in the $\MSbar$ scheme; the mass anomalous dimension at two-loop
order has been determined in Ref.~\cite{Sturm:2009kb} and is also given
in Appendix~\ref{sec:RISMOMAnomalous}. The three-loop result is
presented here and reads, in the Landau gauge in the $\RISMOM$ scheme,%
\footnote{The minus sign on the left-hand side of
  Eqs.~(\ref{eq:gam2m}),~(\ref{eq:gam2mb}) and (\ref{eq:gam2q}) is
  correct, but was missing in the first arXiv version. It is needed to
  be consistent with the definition in Eqs.(\ref{eq:gammam}) and
  (\ref{eq:gammaq}), respectively; see also the note added.}
%
% ========================================================
% = Mass Anomalous dimension RISMOM three loop coefficient =
% ========================================================
%
\begin{eqnarray}
\label{eq:gam2m}
-\gamma_m^{(2),\rismom}\!\!\!\!\!\!&=& 
 - {129\over2}\*\CF^3 
 + \CA^2\*\CF\*\left[
 - {29357\over54} 
 - {5639\*\pi^2\over81}
 - {319\*\pi^3\over486\*\sqrt{3}}
 + {154\*\pi^4\over243} 
 + {55\over3}\*\Sigma
\right. \nonumber\\ &&\left. \hspace{-10mm}
 - {55\over108}\*\ppp{3}
 - {22\*\pi\*\nlog{3}\over3\*\sqrt{3}} 
 + {11\*\pi\*\Log{3}{2}\over18\*\sqrt{3}}
 + {5639\over54}\*\p{3}
 + {176\*\pi^2\over81}\*\p{3}  
\right. \nonumber\\ &&\left. \hspace{-10mm}
 - {44\over27}\*\p{3}^2 
 + 88\*\s{2}{6}
 - 176\*\s{2}{2} 
 - {440\over 3}\*\s{3}{6}
 + {352\over 3}\*\s{3}{2}
 + {220\over3}\*\z3
          \right] 
\nonumber\\ && \hspace{-2cm}  
 + \nf\*\TF\*\left\{\CF^2\*\left[
    77 
 - {152\*\pi^2\over27}
 - {116\*\pi^3\over243\*\sqrt{3}}
 - {464\*\pi^4\over243} 
 - 8\*\Sigma 
 + {8\over27}\*\ppp{3} 
 + {4\*\pi\*\Log{3}{2}\over9\*\sqrt{3}} 
\right.\right. \nonumber\\ &&\left.\left. \hspace{-10mm}
+ {76\over9}\*\p{3}
+ {272\*\pi^2\over81}\*\p{3}
- {68\over27}\*\p{3}^2
- {16\*\pi\*\nlog{3}\over3\*\sqrt{3}}
+ 64\*\s{2}{6} 
- 128\*\s{2}{2} 
\right.\right. \nonumber\\ &&\left.\left. \hspace{-10mm}
- {320 \over 3}\*\s{3}{6} 
+ {256\over 3}\*\s{3}{2}
- {176\over3}\*\z3
                  \right] 
 + \CA\*\CF\* \left[
   {7870\over27} 
 + {3068\*\pi^2\over81} 
 + {58\*\pi^3\over243\*\sqrt{3}} 
 - {56\*\pi^4\over243} 
\right.\right. \nonumber\\ &&\left.\left. \hspace{-10mm}
 - {20\over3}\*\Sigma 
 + {8\*\pi\*\nlog{3}\over3\*\sqrt{3}} 
 - {2\*\pi\*\Log{3}{2}\over9\*\sqrt{3}}
 - {1534\over27}\*\p{3} 
 - {64\*\pi^2\over81}\*\p{3} 
 + {16\over27}\*\p{3}^2
\right.\right. \nonumber\\ &&\left.\left. \hspace{-10mm}
 + {5\over27}\*\ppp{3}
 - 32\*\s{2}{6} 
 + 64\*\s{2}{2} 
 + {160\over3}\*\s{3}{6}
 - {128\over3}\*\s{3}{2}
 + {64\over3}\*\z3
             \right] \right\} 
\nonumber \\&& \hspace{-2.2cm} 
 +\CA\*\CF^2\*\left[
 - 9 
 + {616\*\pi^2\over27 }
 + {319\*\pi^3\over243\*\sqrt{3}} 
 + {1276\*\pi^4\over243} 
 + 22\*\Sigma 
 + {44\*\pi\*\nlog{3}\over3\*\sqrt{3}}
 - {11\*\pi\*\Log{3}{2}\over9\*\sqrt{3}} 
 - {308\over9}\*\p{3} 
\right. \nonumber\\ &&\left. \hspace{-10mm}
 - {748\*\pi^2\over81}\*\p{3}
 + {187\over27}\*\p{3}^2
 - {22\over27}\*\ppp{3}
 - 176\*\s{2}{6} 
 + 352\*\s{2}{2} 
 + {880\over3}\*\s{3}{6} 
\right. \nonumber\\ && \left. \hspace{-10mm}
 - {704\over3}\*\s{3}{2}
 + {88\over3}\*\z3
       \right] 
 + \CF\*\nf^2\*\TF^2\*\left[
 - {856\over27} 
 - {320\*\pi^2\over81} 
 + {160\over 27}\*\p{3}
              \right].
\end{eqnarray}
In the $\RISMOMB$ scheme the two-loop order is again available in
Ref.~\cite{Sturm:2009kb} (see also Appendix~\ref{sec:RISMOMAnomalous})
and the three-loop result in the Landau gauge is given by%
$\mbox{}^\thefootnote$
%
% =================================================================
% = Mass Anomalous dimension RISMOM_\gamma_\mu  three-loop coefficient =
% ==================================================================
\begin{eqnarray}
\label{eq:gam2mb}
-\gamma_m^{(2),\rismomb}&=& 
 - {129\over 2}\*\CF^3 
 + \CF^2\*\CA\*\left[
   {1\over 6 } 
 - {242\*\pi^2\over 27} 
 + {121\over 9 }\*\p{3}
 - {572\*\pi^2\over 81}\*\p{3} 
\right. \nonumber \\&& \left.
 + {143\over 27}\*\p{3}^2
 + 44\*\z3
 + {1232\*\pi^4\over 243} 
 + 22\*\Sigma 
 - {55\over 54 }\*\ppp{3}
               \right]	
\nonumber \\ && \hspace{-2.5cm}
 + \CF\*\CA^2\*\left[
 - {67715\over 108} 
 - {4286\*\pi^2\over 81 }
 + {319\pi^3\over 1944\*\sqrt{3}} 
 + {308\pi^4\over 243} 
 + {55\over 3 }\*\Sigma 
 - {11\over 18}\*\ppp{3} 
\right. \nonumber \\&& \left.
 + {11\*\pi\*\nlog{3}\over 6\*\sqrt{3} }
 - {11\*\pi\*\Log{3}{2}\over 72\*\sqrt{3}} 
 + {2143\over 27}\*\p{3}
 + {88\*\pi^2\over 81}\*\p{3} 
 - {22\over 27 }\*\p{3}^2 
\right. \nonumber \\&& \left.
 -   22\*\s{2}{6} 
 +   44\*\s{2}{2}  
 + {110\over 3}\*\s{3}{6}
 - { 88\over 3}\*\s{3}{2}
 + {308\over 3}\*\z3
              \right]
	\nonumber \\  && \hspace{-2.5cm}
 + \nf\*\TF\*\left\{ 
  \CF^2\*\left[
   {233\over3} 
 + {160\*\pi^2\over 27 }
 - {448\*\pi^4\over 243} 
 - 8\*\Sigma 
 + {10\over 27}\*\ppp{3}
 - {80\over 9 }\*\p{3}
 + {208\*\pi^2\over 81}\*\p{3} 
\right.\right. \nonumber \\ && \left. \left.
 - {52\over 27}\*\p{3}^2
 - 64\*\z3
       \right]	
\right.	\nonumber \\ &&\left. \hspace{-2.5cm}
 + \CA\*\CF\*\left[
   {9395\over 27} 
 + {2576\*\pi^2\over 81} 
 - {29\*\pi^3\over 486\*\sqrt{3} }
 - {112\*\pi^4\over 243} 
 - {20\over 3}\*\Sigma
 + {2\over 9}\*\ppp{3}
 - {2\pi\*\nlog{3}\over 3\*\sqrt{3}} 
\right.\right. \nonumber \\&& \left. \left.
 + {\pi \Log{3}{2}\over 18\*\sqrt{3}} 
 - {1288\over 27}\*\p{3}
 - {32\*\pi^2\over 81}\*\p{3}
 + {8\over 27}\*\p{3}^2
 + 8\*\s{2}{6} 
\right.\right. \nonumber \\&& \left. \left.
 - 16\*\s{2}{2}
 - {40\over 3}\*\s{3}{6}
 + {32\over 3}\*\s{3}{2}
 + {32 \over 3}\*\z3
           \right] 
\right\} 
\nonumber \\&& \hspace{-2.5cm}
 + \nf^2\*\TF^2\*\CF\*\left[
   {160\over 27}\p{3}
 - {1088\over 27} 
 - {320\*\pi^2\over 81 }
  \right].		
\end{eqnarray}
Similarly to the mass anomalous dimension we define the anomalous
dimension of the fermion field as
\begin{equation}
\label{eq:gammaq}
\gamma_q^{x}=2\*{d\log{\Psi_{R,x}}\over d\log{\mu^2}}=
-\gamma^{(0),x}_q\*      \frac{\alpha_s}{4\*\pi}       
-\gamma^{(1),x}_q\*\left(\frac{\alpha_s}{4\*\pi}\right)^2
-\gamma^{(2),x}_q\*\left(\frac{\alpha_s}{4\*\pi}\right)^3
+\mathcal{O}(\alpha_s^4)\,,
\end{equation}
where the renormalized and bare fields are connected by
$\Psi_R=\sqrt{Z_q}\Psi_0$.  The relation between the $\MSbar$ fermion
field anomalous dimension and the ones found in the $\RISMOM$ and
$\RISMOMB$ schemes is, in the Landau gauge, again given by\cite{Chetyrkin:1999pq}
\begin{equation}
\gamma_q^{x}=\gamma_q^{\msbar}-\beta\*
 {\partial C_{q,L}^x\over
  \partial \tfrac{\alpha_s}{4\pi}},
\quad x\in\{\rismom,\;\rismomb, \msbar\}.
\end{equation}
In the $\RISMOM$ scheme $\gamma_q^{\rismom}$ is equal to
$\gamma_q^{\rip}$ to all loop orders in perturbation theory, which has
been shown in Ref.~\cite{Sturm:2009kb}. We have reaffirmed this
equivalence to $\Order{\alpha_s^3}$ by comparing our results with those
of Ref.~\cite{Chetyrkin:1999pq}. For the sake of brevity, we omit this
result since $\gamma_q^{\rip}$, at the three-loop level, can be found in
Ref.~\cite{Chetyrkin:1999pq}. However, such an equality is no longer true
in the $\RISMOMB$ scheme. Therefore, we provide the three-loop
contribution to the fermion field anomalous dimension in this scheme
in the Landau gauge below%
$\mbox{}^\thefootnote$

% ===========================================================
% = Fermion Anomalous Dimension RISMOM three-loop coefficient =
% ===========================================================
\begin{eqnarray}
\label{eq:gam2q}
-\gamma_q^{(2),\rismomb}\!\!\!\!\!\!&=&
 - {3\over2}\*\CF^3
 + \CF^2\*\CA\*\left[ 
   {187\over6 }
 + {286\*\pi^2\over 9} 
 + {319\*\pi^3\over 243\*\sqrt{3}} 
 + {44\*\pi^4\over 243} 
 + {44\*\pi\*\nlog{3}\over 3\*\sqrt{3}}
\right. \nonumber \\&& \left. \hspace{-2cm}
 - {11\*\pi\*\Log{3}{2}\over 9\*\sqrt{3}} 
 - {143\over 3}\*\p{3}
 - {176\*\pi^2\over 81}\*\p{3}
 + {44\over 27}\*\p{3}^2
 + {11\over 54}\*\ppp{3} 
\right. \nonumber \\&& \left. \hspace{-2cm}
 - 176\*\s{2}{6} 
 + 352\*\s{2}{2} 
 + {880\over 3}\*\s{3}{6}
 - {704\over 3}\*\s{3}{2}
 - {80\over3}\*\z3
            \right]
\nonumber \\&&  \hspace{-2.5cm}
 + \CF\*\CA^2\*\left[ 
 - {23933\over 432} 
 - {451\*\pi^2\over 27} 
 - {1595\*\pi^3\over 1944\*\sqrt{3}} 
 - {154\*\pi^4\over 243}
 - {55\*\pi\*\nlog{3}\over 6\*\sqrt{3}} 
 + {55\*\pi\*\Log{3}{2}\over 72\*\sqrt{3}} 
\right. \nonumber \\&& \left. \hspace{-2cm}
 + {451\over 18}\*\p{3}
 + {88\*\pi^2\over 81}\*\p{3}
 - {22\over 27}\*\p{3}^2
 + {11\over 108}\*\ppp{3}
 + 110\*\s{2}{6} 
 - 220\*\s{2}{2} 
\right. \nonumber \\&& \left. \hspace{-2cm}
 - {550\over 3}\*\s{3}{6}
 + {440\over 3}\*\s{3}{2}
 + {31\over 24}\*\z3 
              \right]
\nonumber \\&&  \hspace{-2.5cm}
 + \nf\*\TF\*\left\{\CF^2\*\left[ 
 - {16\over3} 
 - {104\*\pi^2\over 9} 
 - {116\*\pi^3\over 243\*\sqrt{3}} 
 - {16\*\pi^4\over 243 }
 - {16\*\pi\*\nlog{3}\over 3\*\sqrt{3}} 
 + {4\*\pi\*\Log{3}{2}\over 9\*\sqrt{3}} 
 + {52\over 3}\*\p{3}
\right. \right.\nonumber \\&& \left. \left. \hspace{-2cm}
 + {64\*\pi^2\over 81}\*\p{3}
 - {16\over 27 }\*\p{3}^2
 - {2\over 27 }\*\ppp{3}
 + 64\*\s{2}{6} 
 - 128\*\s{2}{2} 
 - {320\over 3}\*\s{3}{6} 
\right. \right.\nonumber \\&& \left. \left. \hspace{-2cm}
 + {256\over 3}\*\s{3}{2} 
 + {16\over3}\*\z3
                        \right] 
\right.\nonumber \\&& \left.\hspace{-2.5cm}
 +\CF\*\CA\*\left[
   {767\over27} 
 + {164\*\pi^2\over 27} 
 + {145\*\pi^3\over 486\*\sqrt{3}} 
 + {56\*\pi^4\over 243} 
 + {10\*\pi\*\nlog{3}\over 3\*\sqrt{3}} 
 - {5\*\pi\*\Log{3}{2}\over 18\*\sqrt{3}}
 - {82\over 9}\*\p{3} 
\right. \right.\nonumber \\ && \left. \left. \hspace{-2cm}
 - {32\*\pi^2\over 81}\*\p{3} 
 + {8\over 27}\*\p{3}^2 
 - {1\over 27}\*\ppp{3} 
 - 40\*\s{2}{6} 
 + 80\*\s{2}{2} 
 + {200\over 3}\*\s{3}{6} 
\right. \right.\nonumber \\&& \left. \left. \hspace{-2cm}
 - {160\over 3}\*\s{3}{2}
 + {8\over 3}\*\z3
          \right]
        \right\}
 - \nf^2\*\CF\*\TF^2\*{80\over27}.
\end{eqnarray}
The one- and two-loop contributions are again given in
Ref.~\cite{Sturm:2009kb} and can also be found in
Appendix~\ref{sec:RISMOMAnomalous}.

The anomalous dimension of the tensor operator is defined by
\begin{equation}
\gamma^x_T={d\log Z^x_T\over d\log\mu^2}
=-\gamma^{(0),x}_T\*      \frac{\alpha_s}{4\*\pi}       
-\gamma^{(1),x}_T\*\left(\frac{\alpha_s}{4\*\pi}\right)^2
-\gamma^{(2),x}_T\*\left(\frac{\alpha_s}{4\*\pi}\right)^3
+\mathcal{O}(\alpha_s^4)\,,
\end{equation}
where the renormalized and bare operators are related by the
renormalization constant $Z^{x}_{\hat{O}}$ with
$\hat{O}_{R}^{x}=Z^{x}_{\hat{O}}\,\hat{O}_{B}$ for $\hat{O}=T$
and $x\in\{\rismom,\;\rismomb\}$. 
The anomalous dimensions can again be obtained to three-loop
order in the Landau gauge by inserting the result for the
matching factors $C^x_{T,L}$ of Eqs.~(\ref{eq:CTRISMOM}) and
(\ref{eq:CTRISMOMB}) into
\begin{equation}
\gamma^x_T=\gamma^{\msbar}_T-\beta{\partial\log C^x_{T,L}\over
  \partial{\alpha_s\over4\pi} }, \quad x\in\{\rismom,\;\rismomb\}.
\end{equation}
The $\MSbar$ anomalous dimension $\gamma^{\msbar}_{T}$ of the tensor
operator is known up to three-loop order and can be taken from Refs.~%
\renewcommand\citeright{,\hspace*{-0.5ex}}%
\cite{Broadhurst:1994se,Gracey:2000am}%
\renewcommand\citeright{]}%
\renewcommand\citeleft{}%
\cite{Gracey:2003yr}%
\renewcommand\citeleft{[}%
. % 
It is also given explicitly in Appendix~\ref{sec:MSbarAnomalous}. The
results for the $\RISMOM$ and $\RISMOMB$ schemes read
\begin{eqnarray}
-\gamma^{(2),\rismom}_T&=&
   \CF^3\*\left[
 - {365\over 6} 
 + 64\*\z3
          \right] 
 + \CA^2\*\CF\*\left[
 - {69607\over 162} 
 - {7693\over 162}\*\p{3} 
\right. \nonumber \\&& \left. \hspace{-2cm}
 + {44\over 81}\*\p{3}^2 
 + {7693\over 243}\*\pi^2 
 - {176\over 243}\*\p{3}\*\pi^2 
 + {770\over 729}\*\pi^4 
 - {11\over 36}\*\ppp{3} 
 + {2816\over 3}\*\s{2}{2} 
\right. \nonumber \\&& \left. \hspace{-2cm}
 - {1408\over 3}\*\s{2}{6} 
 - {5632\over 9}\*\s{3}{2} 
 + {7040\over 9}\*\s{3}{6} 
 + {55\over 9}\*\Sigma 
 + {352\*\nlog{3}\*\pi\over 9\*\sqrt{3}}
\right. \nonumber \\&& \left. \hspace{-2cm}
 - {88\*\Log{3}{2}\*\pi\over 27\*\sqrt{3}}
 + {2552\*\pi^3\over 729\*\sqrt{3}}
 + {1460\over 9}\*\z3 
 \right]
\nonumber \\&&  \hspace{-2.5cm}
+ \CA\*\CF^2\*\left[
   {9883\over 27} 
 + {4004\over 27}\*\p{3} 
 - {121\over 81}\*\p{3}^2 
 - {8008\over 81}\*\pi^2 
 + {484\over 243}\*\p{3}\*\pi^2 
\right. \nonumber \\&& \left. \hspace{-2cm}
 + {572\over 729}\*\pi^4 
 - {44\over 81}\*\ppp{3} 
 - 1760\*\s{2}{2} 
 + 880\*\s{2}{6} 
 + {3520\over 3}\*\s{3}{2} 
\right. \nonumber \\&& \left. \hspace{-2cm}
 - {4400\over 3}\*\s{3}{6} 
 + {22\over 3}\*\Sigma
 - {220\*\nlog{3}\*\pi\over 3\*\sqrt{3}} 
 + {55\*\Log{3}{2}\*\pi\over 9\*\sqrt{3}} 
 - {1595\*\pi^3\over 243\*\sqrt{3}}
 - 200\*\z3
             \right]
\nonumber \\&&  \hspace{-2.5cm}
 + \nf\*\TF\*\left\{
  \CA\*\CF\*\left[
   {17426\over 81} 
 + {890\over 81}\*\p{3} 
 - {16\over 81}\*\p{3}^2 
 - {1780\over 243}\*\pi^2 
 + {64\over 243}\*\p{3}\*\pi^2 
\right.\right. \nonumber \\&& \left.\left. \hspace{-2cm}
 - {280\over 729}\*\pi^4 
 + {1\over 9}\*\ppp{3} 
 - {1024\over 3}\*\s{2}{2} 
 + {512\over 3}\* \s{2}{6} 
 + {2048\over 9}\*\s{3}{2} 
 - {2560\over 9}\*\s{3}{6} 
\right.\right. \nonumber \\&& \left.\left. \hspace{-2cm}
 - {20\over 9}\*\Sigma 
 - {128\*\nlog{3}\*\pi\over 9\*\sqrt{3}}
 + {32\*\Log{3}{2}\*\pi\over 27\*\sqrt{3}} 
 - {928\*\pi^3\over 729\*\sqrt{3}}
 - {256\over 9}\*\z3 
             \right]
\right.\nonumber \\&& \left. \hspace{-2.5cm}
+ \CF^2\*\left[
 - {1883\over 27}
 - {1492\over 27}\*\p{3} 
 + {44\over 81}\*\p{3}^2 
 + {2984\over 81}\*\pi^2 
 - {176\over 243}\*\p{3}\*\pi^2 
 - {208\over 729}\*\pi^4 
\right.\right. \nonumber \\&& \left.\left. \hspace{-2cm}
 + {16\over 81}\*\ppp{3} 
 + 640\*\s{2}{2} 
 - 320\*\s{2}{6} 
 - {1280\over 3}\*\s{3}{2} 
 + {1600\over 3}\*\s{3}{6} 
 - {8\over 3}\*\Sigma 
\right.\right. \nonumber \\&& \left.\left. \hspace{-2cm}
 + {80\*\nlog{3}\*\pi\over 3\*\sqrt{3}}
 - {20\*\Log{3}{2}\*\pi\over 9\*\sqrt{3}}
 + {580\*\pi^3\over 243\*\sqrt{3}}
 + 16\*\z3\right]
      \right\}
\nonumber \\&& \hspace{-2.5cm}
 + \CF\*\nf^2\*\TF^2\*\left[
 - {1784\over 81} 
 + {160\over 81}\*\p{3} 
 - {320\over 243}\*\pi^2
                     \right]
\end{eqnarray}
%%%%%%%%%%%%%%%%%%%%%%%%%%%%%%%%%%%%%%%%%%%%%%
and
%%%%%%%%%%%%%%%%%%%%%%%%%%%%%%%%%%%%%%%%%%%%%%
\begin{eqnarray}
-\gamma^{(2),\rismomb}_T&=&
   \CF^3\*\left[
 - {365\over 6} 
 + 64\*\z3
         \right]
+ \CA^2\*\CF\*\left[
 - {112211\over 324} 
 - {1817\over 81}\*\p{3} 
\right. \nonumber \\&& \left. \hspace{-2cm}
 - {22\over 81}\*\p{3}^2 
 + {3634\over 243}\*\pi^2 
 + {88\over 243}\*\p{3}\*\pi^2 
 + {308\over 729}\*\pi^4 
 - {11\over 54}\*\ppp{3} 
 + {2156\over 3}\*\s{2}{2} 
\right. \nonumber \\&& \left. \hspace{-2cm}
 - {1078\over 3}\*\s{2}{6} 
 - {4312\over 9}\*\s{3}{2} 
 + {5390\over 9}\*\s{3}{6} 
 + {55\over 9}\*\Sigma 
 + {539\*\nlog{3}\*\pi\over 18\*\sqrt{3}}
\right. \nonumber \\&& \left. \hspace{-2cm}
 - {539\*\Log{3}{2}\*\pi\over 216\*\sqrt{3}}
 + {15631\*\pi^3\over 5832\*\sqrt{3}} 
 + {1196\over 9}\*\z3
          \right] 
\nonumber \\&& \hspace{-2.5cm}
 + \CA\*\CF^2\*\left[
   {19271\over 54 }
 + {2717\over 27}\*\p{3} 
 + {11\over 81}\*\p{3}^2 
 - {5434\over 81}\*\pi^2 
 - {44\over 243}\*\p{3}\*\pi^2 
 + {704\over 729}\*\pi^4 
\right. \nonumber \\&& \left. \hspace{-2cm}
 - {55\over 162}\*\ppp{3} 
 - 1408\*  \s{2}{2} 
 + 704\*   \s{2}{6} 
 + {2816\over 3}\*\s{3}{2} 
 - {3520\over 3}\*\s{3}{6} 
 + {22\over 3}\*\Sigma 
\right. \nonumber \\&& \left. \hspace{-2cm}
 - {176\*\nlog{3}\*\pi\over 3\*\sqrt{3}} 
 + {44\*\Log{3}{2}\*\pi\over 9\*\sqrt{3}} 
 - {1276\*\pi^3\over 243\*\sqrt{3}}
 - {644\over 3}\*\z3
            \right]
\nonumber \\&& \hspace{-2.5cm}
 + \nf\*\TF\*\left\{
   \CA\*\CF\*\left[
   {12851\over 81} 
 + {152\over 81}\*\p{3} 
 + {8\over 81}\*\p{3}^2 
 - {304\over 243}\*\pi^2 
 - {32\over 243}\*\p{3}\*\pi^2 
\right.\right. \nonumber \\&& \left.\left. \hspace{-2cm}
 - {112\over 729}\*\pi^4 
 + {2\over 27}\*\ppp{3} 
 - {784\over 3} \*\s{2}{2} 
 + {392\over 3} \*\s{2}{6} 
 + {1568\over 9}\*\s{3}{2} 
 - {1960\over 9}\*\s{3}{6} 
\right.\right. \nonumber \\&& \left.\left. \hspace{-2cm}
 - {20\over 9}\*\Sigma 
 - {98\*\nlog{3}\*\pi\over 9\*\sqrt{3}}
 + {49\*\Log{3}{2}\*\pi\over 54\*\sqrt{3}}
 - {1421\*\pi^3\over 1458\*\sqrt{3}}
 - {160\over 9}\*\z3
             \right] 
\right.\nonumber \\&& \left.\hspace{-2.5cm}
+ \CF^2\*\left[
 - {1901\over 27} 
 - {1024\over 27}\*\p{3} 
 - {4\over 81}\*\p{3}^2 
 + {2048\over 81}\*\pi^2 
 + {16\over 243}\*\p{3}\*\pi^2 
 - {256\over 729}\*\pi^4 
\right.\right. \nonumber \\&& \left.\left. \hspace{-2cm}
 + {10\over 81}\*\ppp{3} 
 + 512   \*\s{2}{2} 
 - 256   \*\s{2}{6} 
 - {1024\over 3}\*\s{3}{2} 
 + {1280\over 3}\*\s{3}{6} 
\right.\right. \nonumber \\&& \left.\left. \hspace{-2cm}
 - {8\over 3}\*\Sigma 
 + {64\*\nlog{3}\*\pi\over 3\*\sqrt{3}} 
 - {16\*\Log{3}{2}\*\pi\over 9\*\sqrt{3}}
 + {464\*\pi^3\over 243\*\sqrt{3}}
 + {64\over 3}\*\z3
        \right]
        \right\}
\nonumber \\&& \hspace{-2.5cm}
 + \CF\*\nf^2\*\TF^2 \*\left[
 - {1088\over 81} 
 + {160\over 81}\*\p{3} 
 - {320\over 243}\*\pi^2
                      \right]
.
\end{eqnarray}
\section{Summary and Conclusion\label{sec:DiscussConclude}}
We have computed the two-loop QCD corrections for matching factors which
convert light quark masses renormalized in the $\RISMOM$ and $\RISMOMB$
schemes to the $\MSbar$ scheme. These schemes are extensions of the
traditional regularization independent momentum-subtraction schemes,
like the $\ARI$ or $\RIp$ scheme. The latter are characterized by an
exceptional subtraction point, whereas in the $\RISMOM$ and $\RISMOMB$
schemes the ultraviolet divergences are subtracted at a symmetric
subtraction point. This allows for a lattice simulation with suppressed
contamination from infrared effects. The perturbative expansion
coefficients at two-loop order are significantly smaller than in the
traditional $\ARI$ and $\RIp$ schemes, with about 0.6\%-0.7\% and 2\% of
the leading order result for the $\RISMOM$ and $\RISMOMB$ schemes for
$\nf=3$ flavors at scales of about 2~GeV.  We also provide the results
for the three-loop anomalous dimensions of the quark field and the mass
in both schemes. The mass anomalous dimension can be used to run the
quark masses to different energy scales.

The use of these matching factors will reduce the uncertainties
associated with the matching procedure for converting the quark mass
from the regularization invariant momentum-subtraction scheme to the
$\MSbar$ scheme due to smaller truncation errors of the perturbative
series. These results will allow for an $\MSbar$ light quark mass
determination only together with lattice simulations in the context of
nonperturbative renormalization, with a significant reduced systematic
error compared to previous determinations which use momentum-subtraction
schemes with an exceptional subtraction point, whose matching factors
show a poor convergence behavior at the required energy scales.

In addition, the availability of both the $\RISMOM$ and $\RISMOMB$
schemes at two-loop order will also allow for a better assessment and
control of the uncertainties. This can be achieved by converting results
derived in the different regularization invariant momentum-subtraction
schemes to the $\MSbar$ scheme, using the above matching factors, and
subsequently comparing and cross-checking the obtained $\MSbar$ results.

Our computation is accomplished by studying at two-loop order quark
bilinear operators inserted into amputated Green's functions for the
vector, axial-vector, scalar, and pseudoscalar operators. We also provide
the corresponding results for the matching factors and the anomalous
dimensions of the tensor operator.

\vspace{2ex}
\noindent
{\bf Acknowledgments:}\\ 
We are grateful to our colleagues of the RBC-UKQCD Collaborations for
many valuable discussions, in particular, to A. Soni for advice and
encouragement. We would like to thank T. Izubuchi and S. Uccirati for
conversations about the master integrals as well as Y. Aoki for
important discussions.  This work was supported by the U.S. DOE under
Contract No. DE-AC02-98CH10886.%\\

\vspace{0.5ex}
\noindent
{\bf Note added:}\\ 
During the finalization of this paper Ref.~\cite{Gorbahn:2010bf} appeared,
where the two-loop QCD corrections of the matching factor
$C^{\rismom}_m$ have also been determined. The result in Eq.~(29) of
Ref.~\cite{Gorbahn:2010bf} is for the symmetric subtraction point in
agreement with our result in Eq.~(\ref{eq:CmLrismom}). In the revised
version the authors of Ref.~\cite{Gorbahn:2010bf} provide additional
results which confirm our results for $C^{\rismomb}_q$ and
$C^{\rismomb}_m$; the corresponding anomalous dimensions agree up to a
global sign which has been rectified in the current version.  We would
like to thank the authors of Ref.~\cite{Gorbahn:2010bf} for
communications and for pointing this out to us.
\begin{appendix}
\section{Master integrals and harmonic polylogarithms\label{sec:MIpolylog}}
The harmonic polylogarithms $H$ of Ref.~\cite{Remiddi:1999ew} are defined
recursively as
\begin{equation}
H(a,a_1,\dots,a_k;x)=\int_{0}^{x}\! dx' f_a(x') H(a_1,\dots,a_k;x')
\end{equation}
with
\begin{equation}
\begin{array}{l@{\qquad}l@{\qquad}l}
\nonumber
f_{1}(x)={1\over1-x},& 
f_{0}(x)={1\over x},&
f_{-1}(x)={1\over1+x}
\end{array}
\end{equation}
and
\begin{equation}
\begin{array}{l@{\qquad}l@{\qquad}l}
\nonumber
H(1;x)=-\nlog{1-x},& 
H(0;x)=\nlog{x},&
H(-1;x)=\nlog{1+x}.
\end{array}
\end{equation}
The constants $\mathcal{H}_{31}^{(2)}$ and $\mathcal{H}_{43}^{(2)}$ in
the master integrals of Eqs.~(\ref{eq:M131}) and (\ref{eq:M243}) can be
expressed in terms of harmonic polylogarithms%
\footnote{The $\vep$ expansion of the master integral
      $\M{1}{3}{1}$ can also be obtained with Refs.~\cite{Davydychev:2000na,Kalmykov:2000qe}.}% 
; their numerical
evaluation is given by%
\footnote{We have checked that our numerical evaluation using
  traditional Feynman parametrizations of these constants agrees with
  the numerically evaluated analytical result.} 
\begin{equation}
\mathcal{H}_{31}^{(2)}= - 6.11477558...,
                       %- 6.1147755821282438728474316305413972504054071260357
\qquad\quad
\mathcal{H}_{43}^{(2)}= +12.45994893....
                       %+12.459948928048829311512150384769877694001078028108
\end{equation}
In the conversion factors these constants always arise in the
form of the sum $\Sigma=\mathcal{H}_{31}^{(2)}+\mathcal{H}_{43}^{(2)}$
in which some of the harmonic polylogarithms cancel. It can be expressed with 
$z_0={i\over\sqrt{3}}$ by
\begin{eqnarray}
\Sigma&=& 	
 {1\over3\*\sqrt{3}}\left[
   12\*\pi\*\z3
 - 12\*\pi\*H(0, +, +; z_0) 
 - 6\*\pi\*H(-, +, +; z_0) 
 + i\*\pi^2\*H(-, +; z_0) 
\right.\nonumber\\&&\qquad\left.
 + 2\*i\*\pi^2\*H(0, +; z_0) 
 - 18\*i\*H(-, +, +, +; z_0) 
 - 36\*i\*H(0, +, +, +; z_0) 
                  \right]
\\ \nonumber
&\simeq&6.34517334592058543866471875422848044360,
\end{eqnarray}
where the functions $H(\pm,\cdots,\pm;x)$ are defined by the following linear
combinations:
\begin{eqnarray}
H(+;x)&=& H(1;x) + H(-1;x), \\
H(-;x)&=& H(1;x) - H(-1;x),\\
H(\pm,a_1,\dots,a_k;x )&=& H(1,a_1,\dots,a_k;x)\pm H(-1,a_1,\dots,a_k;x).
\end{eqnarray}
\section{Gauge dependent parts of the conversion factors \label{sec:Convxi}}
In this section we provide the gauge dependent components of the
conversion factors, where the gauge parameter $\xi$ has been
renormalized in the $\MSbar$ scheme. As mentioned in
Section~\ref{sec:matchfactors}, the fermion field conversion factors in
the $\RISMOM$ scheme are given by the ones in the $\RIp$ scheme;
therefore we omit them here. However, we did check that the gauge dependent
terms were also in agreement with the previous calculations in the $\RIp$
scheme. The results read as follows:
%
% ======================================
% = Cq RISMOM gamma_\mu, xi dependence =
% ======================================
%
\begin{eqnarray}
\label{eq:cqxi}
C^{\rismomb}_{q,\xi}&=&
\left({\alpha_s\over4\*\pi}\right)\*
    \CF\*\xi\*\left[
   {1\over 3}\*\p{3}
 - {3\over 2 }
 - {2\*\pi^2\over 9} 
             \right]
\nonumber\\ 
&+& \left({\alpha_s\over4\*\pi}\right)^2\*\left\{ 
   \CF^2\*\left[ 
   \xi\*\left(
   \Sigma
 - 1 
 + 2\*\z3 
 - {31\*\pi^2\over 9 }
 - {29\*\pi^3\over 162\*\sqrt{3}} 
 + {4\*\pi^4\over 27 } 
\right. \right. \right.\nonumber \\&& \left. \left.\left.
 + 24\*\s{2}{6} 
 - 48\*\s{2}{2} 
 - 40\*\s{3}{6} 
 + 32\*\s{3}{2}
 - {2\*\pi\*\nlog{3}\over \sqrt{3}} 
\right. \right. \right.\nonumber \\ && \left. \left.\left.
 + {\pi\*\Log{3}{2}\over 6\*\sqrt{3}} 
 + {31\over 6 }\*\p{3}
 - {1\over 18 }\*\ppp{3}
       \right)
 + \xi^2\*\left(
  {7\over4 } 
 + {4\*\pi^2\over9} 
 + {4\*\pi^4\over81} 
\right. \right. \right.\nonumber \\&& \left. \left.\left.
 - {2\over3}\*\p{3}
 - {4\*\pi^2\over27}\*\p{3}
 + {1\over9}\*\p{3}^2
        \right)
         \right] 
\right. \nonumber\\ &&\left. \hspace{-2cm}
 + \CA\*\CF\*\left[\xi\*\left(
 - {33\over 4} 
 + {2\*\pi^2\over 9} 
 + {29\*\pi^3\over 432\*\sqrt{3}} 
 + {\pi^4\over 27} 
 + {1\over 2}\*\Sigma 
 + 3\*\z3 
 + {\sqrt{3}\*\pi\*\nlog{3}\over 4} 
\right. \right. \right.\nonumber \\&& \left. \left.\left.
 - {1\over 3 }\*\p{3}
 - {1\over 72}\*\ppp{3}
 - 9 \*\s{2}{6} 
 + 18\*\s{2}{2} 
 - 12\*\s{3}{2}
 + 15\*\s{3}{6} 
\right. \right. \right.\nonumber \\&& \left. \left.\left.
 - {\pi\*\Log{3}{2}\over 16\*\sqrt{3}}
                     \right)
 + \xi^2\*\left(
   {1\over4}\*\p{3} 
 - {3\over 2} 
 - {\pi^2\over 6}  
         \right)
           \right]
     \right\}
 + \Order{\alpha_s^3},
\end{eqnarray}
%
% ===========================
% = Cm RISMOM xi dependence =
% ===========================
%
\begin{eqnarray}
\label{eq:cmxi}
C^{\rismom}_{m,\xi}&=& 
\left({\alpha_s\over4\*\pi} \right)\*
  \CF\*\xi\*\left( 
  {1\over 3}\*\p{3}
 - 1 
 - {2\*\pi^2\over 9} 
           \right)
\nonumber\\
&+&\left({\alpha_s\over4\*\pi} \right)^2\*\left\{ 
   \CF^2\*\left[
   \xi\*\left(
   4 
 + {8\*\pi^2\over 9} 
 + {4\*\pi^4\over 9 }
 + \Sigma 
 - {4\over 3}\*\p{3}
 - {8\*\pi^2\over 9}\*\p{3}
 + {2\over 3}\*\p{3}^2
\right. \right. \right.\nonumber \\&& \left. \left.\left.
 - {1\over 18}\*\ppp{3}
       \right)
 + \xi^2\*\left(
   1 
 + {4\*\pi^2\over 9} 
 + {4\*\pi^4\over 81 }
 - {2\over 3}\*\p{3} 
 - {4\*\pi^2\over 27}\*\p{3} 
 + {1\over 9}\*\p{3}^2
         \right)
         \right]
\right.\nonumber\\ &&\left. \hspace{-1cm}
 + \CA\*\CF\*\left[
  \xi\*\left(
   {7\over 6}\*\p{3}
 - {7\over 2} 
 - {7\*\pi^2\over 9} 
 + {\pi^4\over 27} 
 + {\Sigma\over 2}  
 - {1\over 72}\*\ppp{3}
       \right) 
\right. \right.\nonumber \\ && \left. \left.
 + \xi^2\*\left(
   {1\over 4}\p{3}
 - {3\over 4} 
 - {\pi^2\over 6} 
         \right)
            \right]
\right\} 
+ \Order{\alpha_s^3},
\end{eqnarray}
%
% ======================================
% = Cm RISMOM gamma_\mu, xi dependence =
% ======================================
%
\begin{eqnarray}
\label{eq:cmbxi}
C^{\rismomb}_{m,\xi}&=& 	
 - \left({\alpha_s\over4\*\pi} \right)\*
   {1\over2}\*\CF\*\xi 
\nonumber\\
&+&\left({\alpha_s\over4\*\pi} \right)^2\* \left\{ 
  \CF^2\*\left[
  \xi\*\left(
   2 
 + {26\*\pi^2\over 9} 
 + {29\*\pi^3\over 162\*\sqrt{3}} 
 + {4\*\pi^4\over 27}
 - 2\*\z3 
 + {2\*\pi\*\nlog{3}\over \sqrt{3}}  
\right. \right. \right.\nonumber \\&& \left. \left.\left. 
 - 24\*\s{2}{6} 
 + 48\*\s{2}{2}
 + 40\*\s{3}{6} 
 - 32\*\s{3}{2}
 - {\pi\*\Log{3}{2}\over 6\*\sqrt{3}} 
\right. \right. \right.\nonumber \\&& \left. \left.\left.
 - {13\over 3}\*\p{3}
 - {4\*\pi^2\over 9}\*\p{3}
 + {1\over 3}\*\p{3}^2
       \right) 
 + \xi^2\*\left(
   {1\over 2} 
 + {\pi^2\over 9} 
 - {1\over 6}\*\p{3}
         \right)
        \right]
\right. \nonumber\\ &&\left. \hspace{-1cm}
 + \CA\*\CF\*\left[
 - {3\over8}\xi^2
 + \xi\*\left(
 - {7\over 4}  
 - \pi^2 
 - {29\*\pi^3\over 432\*\sqrt{3}} 
 +  9\*\s{2}{6}
 - 18\*\s{2}{2}
 - 15\*\s{3}{6}
\right. \right. \right.\nonumber \\&& \left. \left.\left.
 + 12\*\s{3}{2}
 - {\sqrt{3}\*\pi\*\nlog{3}\over 4} 
 + {\pi\*\Log{3}{2}\over 16\*\sqrt{3}} 
 + {3\over 2}\p{3} 
       \right)
            \right] 
\right\} 
+ \Order{\alpha_s^3},
\end{eqnarray}
%
% ======================================
% = CT RISMOM xi dependence =
% ======================================
%
\begin{eqnarray}
C^{\rismom}_{T,\xi}&=&
\left({\alpha_s\over4\*\pi} \right)\*\CF\*\xi\*\left(
   {1\over 3} 
 - {1\over 3}\*\p{3} 
 + {2\over 9}\*\pi^2
             \right)
\nonumber\\
&+&\left({\alpha_s\over4\*\pi} \right)^2\*\left\{
   \CF^2\*\xi\*\left[
 - 6\*\p{3} 
 + 4\*\pi^2 
 - {4\over 27}\*\pi^4 
 + {1\over 18}\*\ppp{3} 
 + 64\*   \s{2}{2} 
 - 32\*   \s{2}{6} 
\right. \right.\nonumber \\&& \left. \left.
 - {128\over 3}\*\s{3}{2} 
 + {160\over 3}\*\s{3}{6} 
 - \Sigma 
 + {8\*\nlog{3}\*\pi\over 3\*\sqrt{3}}
 - {2\*\Log{3}{2}\*\pi\over 9\*\sqrt{3}} 
 + {58\*\pi^3\over 243\*\sqrt{3}}
 - {8\over 3}\*\z3
              \right]
\right. \nonumber\\ &&\left. \hspace{-1cm}
 + \CA\*\CF\*\left[
   \xi\*\left(
   {7\over 6} 
 + {5\over 6}\*\p{3} 
 - {5\over 9}\*\pi^2 
 - {1\over 27}\*\pi^4 
 + {1\over 72}\*\ppp{3} 
 - 24\*\s{2}{2} 
\right. \right.\right.\nonumber \\&& \left. \left.\left.
 + 12\*\s{2}{6} 
 + 16\*\s{3}{2} 
 - 20\*\s{3}{6} 
 - {1\over 2}\*\Sigma 
 - {\nlog{3}\*\pi\over \sqrt{3}}
 + {\Log{3}{2}\*\pi\over 12\*\sqrt{3}}
 - {29\*\pi^3\over 324\*\sqrt{3}}
       \right)
\right. \right.\nonumber \\&& \left. \left.
 + \xi^2\*\left(
   {1\over 4} 
 - {1\over 4}\*\p{3} 
 + {1\over 6}\*\pi^2
         \right) 
          \right]
         \right\}
+ \Order{\alpha_s^3},
\end{eqnarray}
%
% ======================================
% = CT RISMOM gamma_\mu, xi dependence =
% ======================================
%
\begin{eqnarray}
C^{\rismomb}_{T,\xi}&=&
 - \left({\alpha_s\over4\*\pi} \right)\*
   {1\over 6}\*
   \CF\*\xi 
\nonumber\\
&+&\left({\alpha_s\over4\*\pi} \right)^2\*
   \left\{
   \CF^2\*\left[
   \xi\*\left(
   1 
 - {16\over 9}\*\p{3} 
 + {1\over 9}\*\p{3}^2 
 + {32\over 27}\*\pi^2 
 - {4\over 27}\*\p{3}\*\pi^2 
\right. \right. \right.\nonumber \\&& \left. \left. \left.
 + {4\over 81}\*\pi^4 
 + 16\*\s{2}{2} 
 - 8\*\s{2}{6} 
 - {32\over 3}\*\s{3}{2} 
 + {40\over 3}\*\s{3}{6} 
 + {2\*\nlog{3}\*\pi\over 3\*\sqrt{3}} 
\right. \right. \right.\nonumber \\&& \left. \left. \left.
 - {\Log{3}{2}\*\pi\over 18\*\sqrt{3}} 
 + {29\*\pi^3\over 486\*\sqrt{3}} 
 - {2\over 3}\*\z3
       \right)
 +  \xi^2\*\left(
   {1\over 12} 
 - {1\over 18}\*\p{3} 
 + {1\over 27}\*\pi^2
        \right)
        \right]
\right. \nonumber\\ &&\left. \hspace{-1cm}
 + \CA\*\CF\*\left[
 - {1\over 8}\*\xi^2
 + \xi\*\left(
 - {7\over 12} 
 + {1\over 2}\*\p{3} 
 - {1\over 3}\*\pi^2 
 - 6\*\s{2}{2} 
 + 3\*\s{2}{6} 
 + 4\*\s{3}{2} 
\right. \right. \right.\nonumber \\&& \left. \left. \left.
 - 5\*\s{3}{6} 
 - {\nlog{3}\*\pi\over 4\*\sqrt{3}} 
 + {\Log{3}{2}\*\pi\over 48\*\sqrt{3}} 
 - {29\*\pi^3\over 1296\*\sqrt{3}}
      \right)
         \right] 
        \right\}
+ \Order{\alpha_s^3}.
\end{eqnarray}
\section{The QCD $\beta$ function and anomalous dimensions in the $\MSbar$ scheme\label{sec:MSbarAnomalous}}
The QCD $\beta$ function is defined by
\begin{equation}
\label{eq:beta}
\beta={d\tfrac{\alpha_s}{4\*\pi}\over d\log{\mu^2}}=
-\beta_0\*\left(\frac{\alpha_s}{4\*\pi}\right)^2
-\beta_1\*\left(\frac{\alpha_s}{4\*\pi}\right)^3
+\mathcal{O}\left(\alpha_s^4\right)
\end{equation}
and known up to four-loop order~\cite{Gross:1973id,Politzer:1973fx,tHooft:1972,Caswell:1974gg,Jones:1974mm,Egorian:1978zx,Tarasov:1980au,Larin:1993tp,vanRitbergen:1997va,Czakon:2004bu}. The lowest two orders are given by
\begin{eqnarray}
\label{eq:beta0}
\beta_0&=&{11\over3}\*\CA-{4\over3}\*\TF\*\nf,\\
\label{eq:beta1}
\beta_1&=&{34\over3}\*\CA^2-4\*\CF\*\TF\*\nf-{20\over3}\*\CA\*\TF\*\nf.
\end{eqnarray}
The $\MSbar$ mass anomalous 
\renewcommand\citeright{,\hspace*{-0.5ex}}%
dimension~\cite{Tarrach:1980up,Tarasov:1982aa}%
\renewcommand\citeright{]}%
\renewcommand\citeleft{}%
\cite{Larin:1993tq,Chetyrkin:1997dh,Vermaseren:1997fq}
\renewcommand\citeleft{[}%
is defined in Eq.~(\ref{eq:gammam}) and reads up to order~$\alpha_s^3$
\begin{eqnarray}
\gamma_m^{(0),\msbar}&=&3\*\CF,\\
\gamma_m^{(1),\msbar}&=&{3\over2}\*\CF^2+{97\over6}\*\CF\*\CA-{10\over3}\*\CF\*\TF\*\nf,\\
\gamma_m^{(2),\msbar}&=&
   {129\over2}\*\CF^3
 - {129\over4}\*\CF^2\*\CA
 + {11413\over108}\*\CF\*\CA^2
 + \CF^2\*\TF\*\nf\*\left(48\*\z3-46\right)
\nonumber\\&&
 - \CF\*\CA\*\TF\*\nf\*\left(48\*\z3+{556\over27}\right)
 - {140\over27}\*\CF\*\TF^2\*\nf^2.
\end{eqnarray}
The lowest orders of the $\MSbar$ anomalous dimension of the fermion
field~\cite{Chetyrkin:1999pq,Gracey:2003yr} defined in
Eq.~(\ref{eq:gammaq}) for $\xi=0$ (Landau gauge) are given by\\
\begin{eqnarray}
\gamma_q^{(0),\msbar}&=&
0, \qquad%\\
\gamma_q^{(1),\msbar} =
 - \CF^2\*{3\over2}
 + \CF\*\CA\*
    {25\over4} 
 - 2\*\CF\*\TF\*\nf,\\
\label{eq:gammaq2}
\gamma_q^{(2),\msbar}&=&
   \CF^3\*{3\over2}
 - \CA\*\CF^2\*\left({143\over4} - 12\*\z3\right) 
 + \CA^2\*\CF\*\left(
      {9155\over 144 }
    - {69\over8}\*\z3 
           \right)
\nonumber\\&&
 + \nf\*\TF\*\left(
   3\*\CF^2 
 - \CA\*\CF\*
   {287\over9} 
             \right)
 + \CF\*\nf^2\*\TF^2\*{20\over9}.
\end{eqnarray}
The Casimir operators in Eqs.~(\ref{eq:beta0})-(\ref{eq:gammaq2}) of the
SU($N_c$) group are given by $\CA=N_c$, $\CF=(N_c^2-1)/(2\*N_c)$, where
$N_c$ is the number of colors.\\
For our calculation of the tensor anomalous dimensions in the $\RISMOM$
and $\RISMOMB$ schemes, we need the corresponding $\MSbar$ anomalous 
dimension of the tensor operator up to three-loop order 
\renewcommand\citeright{,\hspace*{-0.5ex}}%
\cite{Broadhurst:1994se,Gracey:2000am}%
\renewcommand\citeright{]}%
\renewcommand\citeleft{}%
\cite{Gracey:2003yr}%
\renewcommand\citeleft{[}, %
which we have taken from these references. The first three orders read
\begin{eqnarray}
\gamma^{(0),\msbar}_T&=&
   \CF, \qquad%\\
\gamma^{(1),\msbar}_T=
 -  \CF^2\*{19\over2}
 + \CF\*\CA\*{257\over18} 
 - \CF\*\nf\*\TF\*{26\over9},\\
%%%%%
\gamma^{(2),\msbar}_T&=&
 -  \CF^3\*\left( 64\*\z3 - {365\over 6} \right)
 + \CA\*\CF^2\*\left( 112\*\z3 -  {6823\over 36} \right)
 - \CA^2\*\CF\*\left( 40\*\z3 - {13639\over 108} \right)
\nonumber\\&&
 + \nf\*\TF\*\left[ \CF^2\*\left( 16\*\z3 + {98\over 9} \right)
 -                  \CA\*\CF\*\left( 16\*\z3 + {1004\over 27} \right)
            \right]
 - {4\over 3}      \*\CF\*\nf^2\*\TF^2 
.
\end{eqnarray}

\section{The one- and two-loop anomalous dimensions in the $\RISMOM_{(\gamma_{\mu})}$ scheme\label{sec:RISMOMAnomalous}}
The one- and two-loop anomalous dimensions have already been determined
in Ref.~\cite{Sturm:2009kb}. In order to avoid conventional subtleties
we also provide these results here for completeness. The one-loop
results in the Landau gauge are given by
\begin{eqnarray}
\gamma_m^{(0),\rismom}&=&3\*\CF=\gamma_m^{(0),\rismomb},
\qquad
\gamma_q^{(0),\rismomb}=0,\\
\gamma_T^{(0),\rismom}&=&\CF=\gamma_T^{(0),\rismomb}.
\end{eqnarray}
The two-loop results in the Landau gauge read
\begin{eqnarray}
-\gamma_m^{(1),\rismom}&=&
 - \CF^2\*{3\over2}
 - \CA\*\CF\*\left(
   {185\over6} 
 + {22\over9}\*\pi^2 
 - {11\over3}\*\p{3}
            \right)
\nonumber\\
&-&\CF\*\TF\*\nf\*\left( 
   {4\over3}\*\p{3} 
 - {26\over3} 
 - {8\over9}\*\pi^2 
                   \right),\\
-\gamma_m^{(1),\rismomb}&=&
 - \CF^2\*{3\over2}
 - \CA\*\CF\*\left(
   {69\over2} 
 + {22\over9}\*\pi^2 
 - {11\over3}\*\p{3}
            \right)
\nonumber\\
&-&\CF\*\TF\*\nf\*\left( 
   {4\over3}\*\p{3} 
 - 10 
 - {8\over9}\*\pi^2
                 \right),
\\
-\gamma_q^{(1),\rismomb}&=&
   \CF^2 \*{3\over2}
 - \CA\*\CF\*{31\over12}
 + \CF\*\TF\*\nf\*{2\over3},
\end{eqnarray}
%%%%%%%%%%%%%%%%%%%%%%%%
\begin{eqnarray}
-\gamma^{(1),\rismom}_T&=&
   \CF^2\*{19\over 2} 
 - \CA\*\CF\*\left(
   {115\over 6} 
 - {11\over 9}\*\p{3} 
 + {22\over 27}\*\pi^2
             \right) 
\nonumber\\
&+&\CF\*\nf\*\TF\*\left(
   {14\over 3} 
 - {4\over 9}\*\p{3} 
 + {8\over 27}\*\pi^2
                 \right),\\
%%%%%%%%%%
-\gamma^{(1),\rismomb}_T&=&
   \CF^2 \*{19\over 2}
 - \CA\*\CF\*\left(
   {31\over 2} 
 - {11\over 9}\*\p{3} 
 + {22\over 27}\*\pi^2
            \right) 
\nonumber\\
&+&\CF\*\nf\*\TF\*\left(
   {10\over 3} 
 - {4\over 9}\*\p{3} 
 + {8\over 27}\*\pi^2
                 \right).
\end{eqnarray}
%

%%%%%%%%%%%%%%%%%%%%%%%%%%%%%%%%%%%%%%%%%%%%%%%%%%%%%%%%%%%%%%%%%%%%%%%%%%

%
%
%
%
%
%
\end{appendix}

\end{document}